\documentclass[12pt, draftclsnofoot, onecolumn]{IEEEtran}
\usepackage{amsthm,amsmath,amssymb}
\usepackage{mathrsfs}
\usepackage{amsfonts,amssymb}
\usepackage[T1]{fontenc}
\usepackage{hyperref}
\usepackage{multirow}
\usepackage{diagbox}
\usepackage{cite}
\usepackage[pdftex]{graphicx}
\usepackage{graphicx, subfigure}
\usepackage{amsmath}
\usepackage[cmintegrals]{newtxmath}
\usepackage{threeparttable}
\begin{document}
\newcommand{\tabincell}[2]{\begin{tabular}{@{}#1@{}}#2\end{tabular}}

\title{CAnet: Uplink-aided Downlink Channel Acquisition in FDD Massive MIMO using Deep Learning}

\author{\normalsize {Jiajia~Guo, 
Chao-Kai~Wen, \IEEEmembership{\normalsize {Senior Member,~IEEE}},
~Shi~Jin, \IEEEmembership{\normalsize {Senior Member,~IEEE}},
%
}

\thanks{J.~Guo and S.~Jin are with the National Mobile Communications Research
Laboratory, Southeast University, Nanjing, 210096, P. R. China (email: jiajiaguo@seu.edu.cn, jinshi@seu.edu.cn).}
\thanks{C.-K.~Wen is with the Institute of Communications Engineering, National Sun Yat-sen University, Kaohsiung 80424, Taiwan (e-mail: chaokai.wen@mail.nsysu.edu.tw).}
%
}

\maketitle

\begin{abstract}

In frequency-division duplexing systems, the downlink channel state information (CSI) acquisition scheme leads to high training and feedback overheads.
In this paper, we propose an uplink-aided downlink channel acquisition framework using deep learning to reduce these overheads.
Unlike most existing works that focus only on channel estimation or feedback modules, to the best of our knowledge, this is the first study that considers the entire downlink CSI acquisition process, including downlink pilot design, channel estimation, and feedback. First, we propose an adaptive pilot design module by exploiting the correlation in magnitude among bidirectional channels in the angular domain to improve channel estimation.
Next, to avoid the bit allocation problem during the feedback module, we concatenate the complex channel and embed the uplink channel magnitude to the channel reconstruction at the base station.
Lastly, we combine the above two modules and compare two popular downlink channel acquisition frameworks.
The former framework estimates and feeds back the channel at the user equipment subsequently.
The user equipment in the latter one directly feeds back the received pilot signals to the base station.
Our results reveal that, with the help of uplink,  directly feeding back the pilot signals can save approximately 20\% of feedback bits, which provides a guideline for future research.
\end{abstract}


\IEEEpeerreviewmaketitle

\section{Introduction}
\label{introduction}

\IEEEPARstart{S}{ince} the standardization of the fifth generation (5G) communication system has gradually been solidified, researchers in the communication community are beginning to turn their attention to 5G evolution and 6G \cite{9170653}.
Further advancement, such as massive multiple-input and multiple-output (MIMO) with increased antennas, distributed antenna arrangement combined with new network topology, and increased layers for spatial multiplexing, is expected \cite{docomo2020white}.
A massive MIMO architecture is integral to 5G networks, especially as a key technology to utilize millimeter waves effectively \cite{6824752,6798744}.
In massive MIMO systems, base station (BSs) are equipped with a large number of antennas to improve spectral and energy efficiencies through relatively simple (linear) processing.
However, reaping the full benefits of massive MIMO requires perfect knowledge of the
downlink channel state information (CSI) at the BSs.

In time-division duplexing (TDD) systems, the downlink CSI can be inferred from the uplink CSI on the basis of the assumption of channel reciprocity.
In frequency-division duplexing (FDD) systems, where the downlink and the uplink use different frequency bands, channel reciprocity no longer holds.
To obtain the downlink CSI at the BS, a two-stage procedure is required.
First, the BS transmits pilot symbols to the user equipment (UE).
Then, the UE feeds back the estimated channel to the BS through the uplink.
This CSI acquisition strategy leads to considerable capacity reduction.
To simplify the channel estimation process, the pilots are designed to be orthogonal; hence, the number of pilots is proportional to the antenna number at the BS.
The feedback overhead is also proportional to the number of antennas at the BS \cite{7475896}.
With the large antenna dimensions of massive MIMO, the overhead of training and feedback is prohibitively large and consumes the entire system capacity, thereby hindering the implementation of FDD systems in future communications, especially in an uplink centric broadband communication scenario that will enable a 10-fold increase in uplink bandwidth \cite{55g}.
Nevertheless, compared with TDD, the FDD mode is more effective for the systems that have symmetric traffic and delay-sensitive applications \cite{7373680}.

Several possible ways have been proposed to enable FDD in massive MIMO systems by reducing the downlink CSI acquisition overhead.
Compressed sensing (CS), which is based on the sparsity assumption under some basis or dictionary, is regarded as a potential technology to alleviate this problem \cite{8350399}.
In massive MIMO systems, channel responses corresponding to different antennas are correlated, and a hidden joint sparsity property from shared local scatterers exists in the signal propagation environment.
Research has shown that this sparsity can be exploited to shrink the channel to an effective one that has a much lower dimension.
For example, the authors in \cite{7373680} propose CS-based pilot design schemes to reduce the training overhead by exploiting the hidden joint sparsity.
For the downlink CSI feedback, CS algorithms can be exploited to compress and reconstruct CSI at the UE and the BS, respectively.
In this way, the feedback overhead can be greatly reduced, as in \cite{7434506}.
However, the CS-based pilot design, channel estimation, and CSI feedback only exploit the sparsity as the prior information and ignore the fact that the CSI is highly related to the environment of the UE.
Meanwhile, CS problems are often solved using iterative algorithms, which are of high complexity and are time consuming.
These drawbacks hinder the implementation of CS-based algorithms in communications. 

Recently, deep learning (DL) has achieved great success in many applications, such as computer vision and natural language processing. 
In communications, among all technological works pertaining to 6G, DL has been regarded as one of the most eye-catching ideas, which will be the ground-breaking technology in 6G \cite{9170653,docomo2020white}.
Considerable DL-based applications to communications exist, especially in the physical layer \cite{8663966,8233654}.
DL is widely used to enhance certain modules of conventional communication systems, such as channel prediction \cite{8795533,9277535}, joint channel estimation and signal detection \cite{8052521,jiang2018artificial}, and beamforming (BF) design \cite{8935405}. 
Moreover, an end-to-end communication has been proposed to change the classic communication structure completely \cite{8985539}.

The CsiNet introduced by \cite{8322184} is the first attempt to apply DL to CSI feedback.
CsiNet is based on autoencoder architecture in which the encoder and the decoder compress and reconstruct the downlink CSI, respectively.
Different from the CS-based algorithms that exploit only the sparsity characteristic and solve optimization problems by using iterative methods with high complexity, the DL-based method can learn the environment and realize reconstruction with high accuracy in a short time by exploiting graphics processing units (GPUs) \cite{8663966}.
The temporal correlation is exploited using long short-term memory (LSTM) architecture to improve the feedback accuracy of time-varying channels in \cite{8482358}.
On the basis of the high correlation in magnitude among bidirectional channels in the delay domain, DualNet in \cite{8638509} uses the uplink magnitude information to help in the reconstruction of the downlink channel magnitude.
The authors in \cite{guo2020dl,yang2020distributed} introduce CoCsiNet and distributed DeepCMC, which exploit the correlation among nearby UE.
A quantization module and a multiple-rate feedback framework are introduced to the DL-based CSI feedback in \cite{8972904}.
To reduce the feedback bits further, the authors in \cite{yang2020distributed} add an entropy-coding module following the quantization.
The authors in \cite{guoJSAC} consider the BF design with DL-based CSI feedback and propose joint feedback and BF design frameworks for single- and multi-cell systems, respectively.
The neural network (NN) complexity, feedback noise, joint channel estimation and feedback, and feedback safety are studied in \cite{9136588,lu2020binary,9076084,chenTVT,9143575}, respectively.
The performance of DL-based CSI feedback in practical systems is evaluated in \cite{9200894,8117578}.
The experimental results in \cite{8117578} show that the DL-based
channel feedback framework can offer an average of 73\% airtime
overhead reduction and increase throughput by approximately 69\%
compared with the 802.11 feedback protocols.
These results show the high potential of DL-based CSI feedback in the 6G era.

\begin{figure*}[t]
    \centering 
    \includegraphics[scale=0.5]{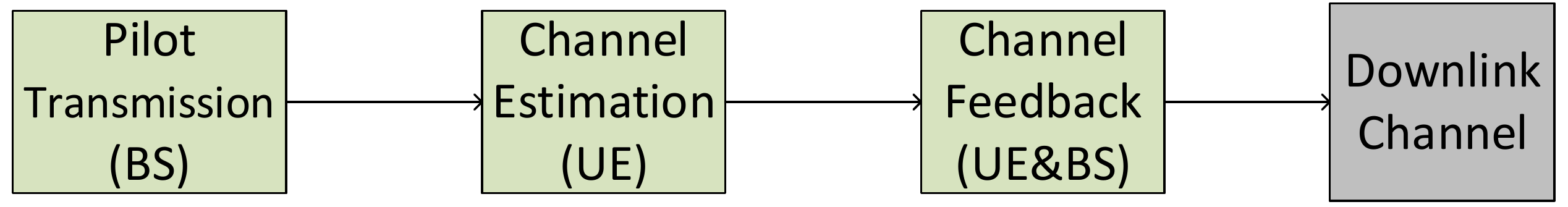}
    \caption{\label{CSIacq}Diagram of downlink CSI acquisition in FDD massive MIMO systems. It mainly consists of three steps: pilot transmission, channel estimation, and channel feedback.}  
\end{figure*}

Most of the aforementioned works, except \cite{chenTVT}, consider only the feedback process and assume that the UE has obtained perfect downlink CSI, which is impossible in practical systems.
As shown in Fig. \ref{CSIacq}, the downlink CSI acquisition of FDD systems comprises three main steps: pilot transmission, channel estimation, and feedback.
These steps interact and affect one another.
Even if one step achieves optimal performance, the performance of the total system may not be optimal.
For example, given that the feedback is lossy and introduces errors into the CSI, whether the downlink channel should be estimated as accurately as possible at the expense of the high training overhead causes a doubt.
Therefore, a complete downlink CSI acquisition framework for FDD massive MIMO systems should be designed.

In this paper, we propose a DL-based uplink-aided downlink \underline{C}SI \underline{a}cquisition framework (called CAnet) for FDD massive MIMO systems.
If the duplex distance is not large in FDD systems, a loose and abstract form of reciprocity exists between the downlink and uplink channels \cite{hugl2002spatial,molisch2003geometry,8383706,8354789,8697125}.
For instance, assuming that the angle of arrival of the signals in the uplink transmission is the same as the angle of departure (AoD) of the signals in the downlink transmission is reasonable.
That is, the directions of signal paths are invariant to carrier frequency shift \cite{8383706}.
This characteristic is exploited in this study to alleviate the acquisition overhead of the downlink CSI, including channel estimation and feedback.

In existing wireless systems, the downlink CSI is estimated at the UE from the pilot signals sent by the BS.
During this phase, the pilot design exploits only the statistical CSI information, and the pilot is fixed.
In other words, the pilot cannot be adjusted in accordance with the real channel environment.
The BS has the knowledge of the uplink CSI; accordingly, designing the downlink pilot by exploiting the obtained downlink CSI is rational.
Toward this end, the pilot design and channel estimation modules in this paper are jointly trained via an end-to-end learning.
As such,  we introduce an adaptive pilot design strategy, where  
the downlink pilot can adapt to the corresponding uplink channel rather than
the widely used statistical channel information. Compared with the fixed pilot design  \cite{8861085},  such an adaptive strategy can achieve considerable performance gains.

In existing DL-based feedback works, to exploit the correlation between the bidirectional channels in the delay domain \cite{8638509} and the nearby UE in the angular domain \cite{guo2020dl}, the magnitude and the phase of the downlink are fed back separately, which leads to a complicated bit allocation problem between the feedback of CSI magnitude and phase.
To avoid this problem, we propose to send back the complex CSI directly and fuse the side information, i.e., downlink CSI magnitude, into the decoder.
Compared with the traditional algorithms that need hand-designed information mixture, the DL-based method can fuse the information automatically \cite{9277535}.
In this work, NNs learn and exploit the correlation through an end-to-end learning.

Practical wireless systems first estimate the downlink channel, then compress and feed back it to the BS.
Whether estimating the downlink channel at the UE first, which needs extra computing power compared with directly feeding back the received signals, is necessary is being questioned.
In this process, additional side information, such as the uplink CSI, is available at the BS and can help reconstruct the downlink channel.
Nonetheless, this information is unavailable at UE in real applications.
If the UE has to estimate the full CSI, then the received pilot signals have to contain all information about the downlink channel, which necessitates increased training and feedback overheads.
If the UE directly feeds back the pilot signals, then the BS reconstructs the downlink CSI by exploiting the feedback pilot signals and side information, i.e., uplink channel, which may reduce the CSI acquisition overhead.
Thus, we propose and compare two different feedback strategies: explicit and implicit feedback. 
In the former, we divide the downlink CSI acquisition into two steps: joint uplink-aided pilot design and channel estimation and uplink-aided feedback.
The implicit feedback considers the downlink CSI acquisition as an indivisible task.
Once the UE obtains the pilot signal, it directly compresses and quantizes the received signal and then feeds back the bitstreams to the BS.
The entire process is realized using an NN-based module rather than two separated modules.
If the proposed NNs are applied to practical systems, the complexity problem cannot be ignored.
Consequently, we apply NN weight pruning, a representative NN compression technique, to the proposed 
CAnet framework, thereby greatly reducing the requirement of memory, power consumption, and model storage space.

The major contributions of this paper are summarized as follows:
\begin{itemize}
\item 
To design the pilot in accordance with the instantaneous channel characteristic rather than fixed pilot in existing systems, we propose an adaptive pilot design strategy by exploiting the correlation of the bidirectional channels in the magnitude domain, thereby reducing channel estimation errors.

\item On the basis of the DL's ability to mix information automatedly, we directly feed back the complex downlink channel and embed the uplink channel magnitude to the reconstruction module at the BS rather than feeding back the magnitude and the phase separately, thereby avoiding the complicated bit allocation problem that hinders the deployment of the existing uplink-aided CSI feedback.

\item 
Existing systems estimate the downlink CSI first at the UE, which may be unnecessary.
Therefore, we propose and compare two types of DL-based CSI acquisition frameworks. The first framework initially designs downlink pilots and realizes channel estimation and then feeds back the estimated CSI by using NNs.
The second one regards the entire downlink CSI acquisition process as an indivisible task and realizes it by using an NN-based module, which is an implicit feedback strategy. 
The simulation results show that, with the aid of uplink, directly feeding back the pilot signals can save approximately 20\% of feedback bits.
\item We also consider the high complexity of the NN-based CAnet framework and apply an NN compression technique, i.e., NN weight pruning, to the proposed NNs.
The weight pruning greatly reduces the NN complexity without performance drop.
\end{itemize}

The rest of this paper is organized as follows. 
Section \ref{s2} introduces the massive MIMO channel model and the conventional entire downlink CSI acquisition process. 
Sections \ref{s3} and \ref{s4} present the motivations and NN details of the uplink-aided downlink pilot design and channel estimation framework and the uplink-aided downlink CSI feedback framework, respectively.
Section \ref{s5} introduces two CSI acquisition frameworks, CAnet-J and CAnet-S, and applies NN weight pruning to the proposed NNs.
Section \ref{s6} provides the numerical results of the proposed NN frameworks and demonstrates the effects of NN weight pruning on the proposed NNs.
Section \ref{s7} concludes our paper.

Notations used in this paper are as follows.
Upper and lower boldface letters are used to denote matrices and vectors, respectively.
$(\cdot)^*$, $(\cdot)^T$, and $(\cdot)^H$ represent the complex conjugate, transpose, and Hermitian transpose, respectively.

\section{System model}
\label{s2}
After introducing the physical channel model adopted in this study, we will describe the downlink CSI acquisition process in FDD massive MIMO systems.
\subsection{Physical channel model}
 \label{2A}
\begin{figure*}[t]
    \centering 
    \includegraphics[scale=0.4]{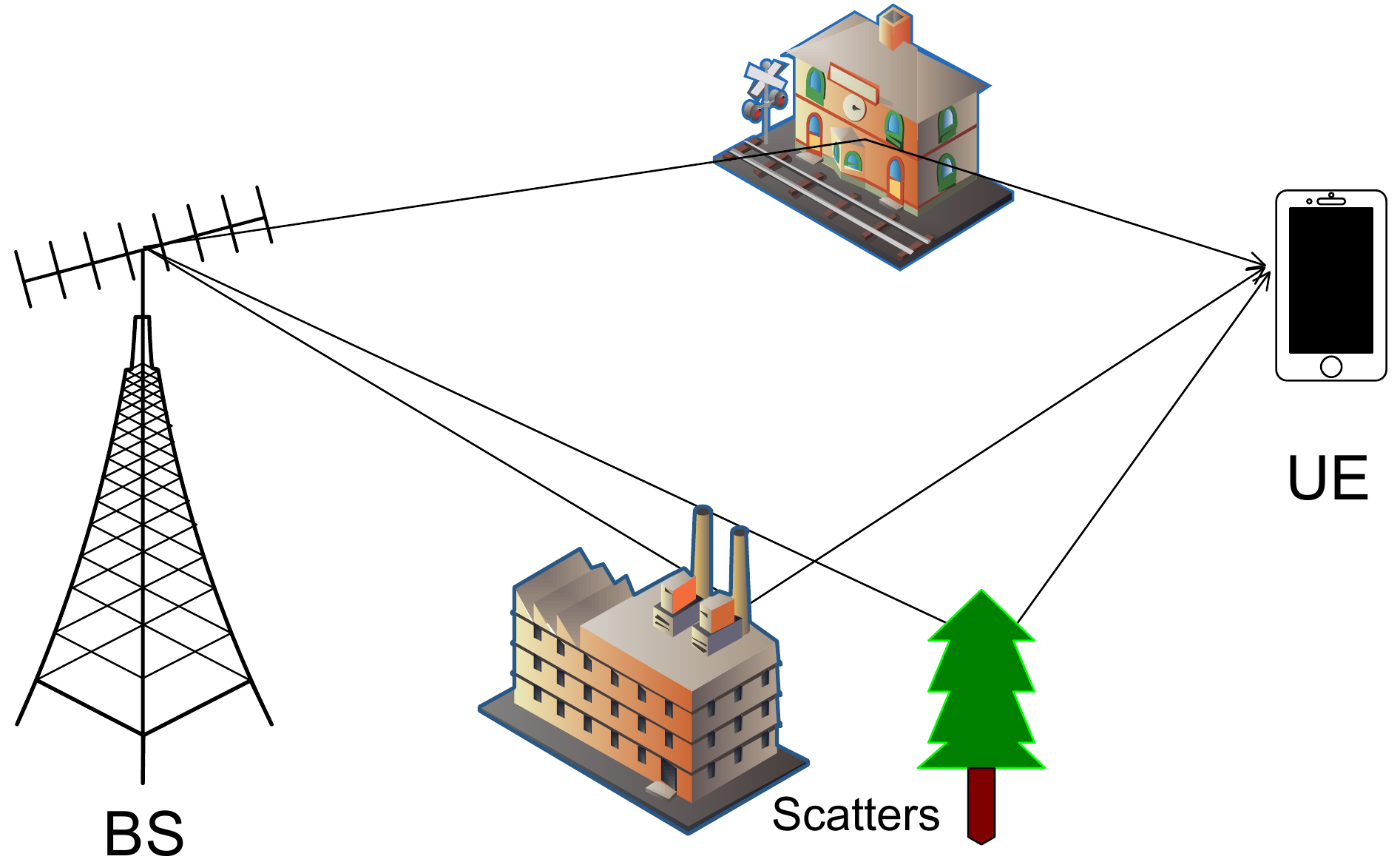}
    \caption{\label{FiniteDimensionalChannel}Illustration of signal propagation in a typical massive MIMO system.}  
\end{figure*}
In this paper, as shown in Fig. \ref{FiniteDimensionalChannel}, we consider a narrowband massive MIMO system, where a uniform linear array (ULA) is equipped at the BS with $N_{\rm t}\gg 1$ antennas and the UE is equipped with a single antenna. 
We adopt the narrowband model here for a simple illustration; nevertheless, the following analysis and the proposed approach are not restricted to this setting and are applicable for wideband systems. 
If wideband systems are adopted, one more channel dimension will be added.
The channel can be regarded as an image rather than a vector in this paper, and CsiNet-based NNs can be used.
In this paper, we adopt the spatial channel model (SCM) \cite{3gpp996}, where the channel ${\bf{h}}_{\rm s} \in \mathbb{C}^{N_{\rm t} \times  1}$ between the BS and the UE can be formulated as
\begin{equation}
{\bf{h}}_{\rm s} =  \sum_{c=1}^{N_c}\sum_{s=1}^{N_s}\xi_{c,s} {\bf{a}}(\theta_{c,s}) ,
\end{equation}
where $N_c$ and $N_s$ represent the number of scattering clusters and the number of subpaths in each scattering cluster, respectively; $\xi_{c,s}$ denotes the complex gain of the $s$-th subpath in the $c$-th cluster; $\theta_{c,s}$ is the corresponding AoD; and ${\bf{a}}(\cdot)$ is the steering vector.
When the BS is equipped with a ULA, the steering vector ${\bf{a}}(\theta)$ can be written as
\begin{equation}
{\bf{a}}(\theta) = [1, e^{-j2\pi \frac{d}{\lambda}\sin(\theta)},\cdots  ,e^{-j2\pi \frac{(N_{\rm t}-1)d}{\lambda}\sin(\theta)}]^T,
\end{equation}
where $\lambda$ and $d$ stand for the wavelength of the downlink and the distance between adjacent antennas at the BS, respectively.

Given that the uplink and the downlink of a certain UE share common physical paths and similar spatial propagation characteristics, a correlation exists among bidirectional channels when the duplex distance is not large \cite{8354789}.
For example, in \cite{8697125}, the downlink CSI is constructed by exploiting frequency-independent parameters between the uplink and downlink channels in the angular-space channel.
The authors in \cite{8383706} claim that the bidirectional channels show the common spatial structure in the angular domain.
A joint uplink/downlink channel estimation algorithm is introduced to obtain accurate bidirectional channels.  
Therefore, exploiting uplink CSI magnitude in the angular domain to design downlink pilot and aid feedback in FDD massive MIMO systems is rational.

\subsection{Downlink CSI acquisition process}
\label{2B}
As shown in Fig. \ref{CSIacq}, typical downlink CSI acquisition consists of pilot transmission, channel estimation, and feedback.
During channel estimation, the received signal $\bf y$ at the UE in $M$ successive time slots can be expressed as
\begin{equation}
\label{eqP}
{\bf y} ={\bf h}^T _{\rm s} {\bf X} + {\bf n} ,
\end{equation}
where ${\bf{X}}\in {\mathbb C}^{N_{\rm t} \times M}$ denotes the downlink pilot signals after precoding transmitted by the BS and  ${\bf n}\in {\mathbb C}^{1 \times M}$ is the complex additive white Gaussian noise with zero mean and variance $\sigma^2$, which is independent across $M$ time slots.
To utilize the angular-domain sparsity in massive MIMO systems, we define the channel in the angular domain as
\begin{equation}
{\bf{h}}_{\rm a} = {\bf F} {\bf h}_{\rm s} ,
\end{equation}
where ${\bf F} $ is an $N_{\rm t} \times N_{\rm t}$ discrete Fourier transform matrix.
Thus, (\ref{eqP}) can be rewritten as
\begin{equation}
\label{transmission}
{\bf y}  =({\bf F}^* {\bf h}_{\rm a})^T {\bf X} + {\bf n} =  {\bf h}_{\rm a}^T {\bf F}^H{\bf X} + {\bf n} =  {\bf h}_{\rm a}^T  \widetilde{X}+ {\bf n} =(\widetilde{X}^T  {\bf h}_{\rm a}+ {\bf n}^T )^T,
\end{equation}
where $\widetilde{X} ={\bf F}^H{\bf X}$.
During the downlink estimation, the UE estimates the channel on the basis
of the obtained 
$\bf y$ and $\widetilde{X}$.

\begin{figure*}[t]
    \centering 
    \includegraphics[scale=0.85]{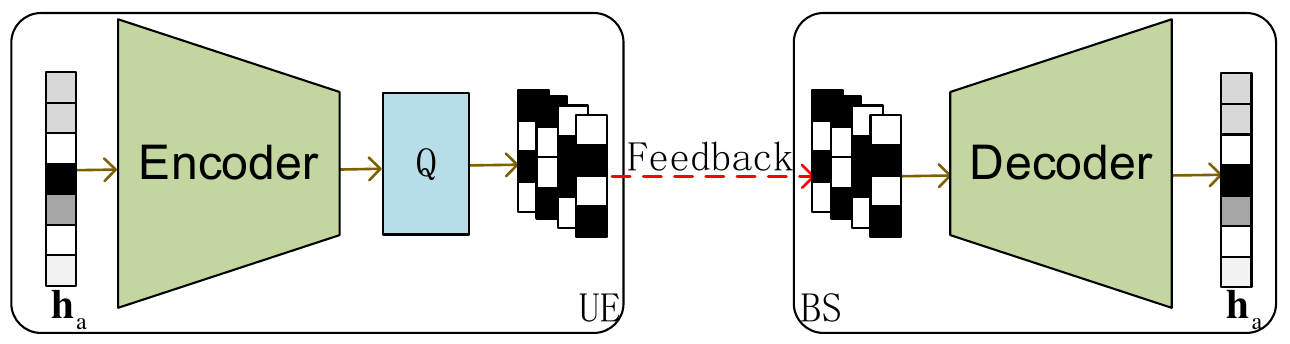}
    \caption{\label{autoencoder}Illustration of the DL-based CSI feedback framework, where the encoder at the UE compresses downlink CSI and the decoder at the BS  reconstructs downlink CSI.}  
\end{figure*}

As shown in Fig. \ref{autoencoder}, during the DL-based CSI feedback phase, the compression and reconstruction parts are realized using the encoder at the UE and the decoder at the BS, respectively.
At the encoder, the dimension of channel ${\bf h}_{\rm a}$ is reduced by decreasing the neuron number of the FC layer.
A quantization module, which discretizes the compressed CSI and generates bitstreams for uplink feedback, follows the encoder \footnote{
Since the entropy coding module after quantization and the dequantization module at the decoder are lossless, we ignore them in this paper.}.
Once the BS obtains the feedback measurements, the decoder acquires an initial coarse downlink channel by increasing the neuron number and then stacks additional NN layer to refine the coarse channel, such as the RefineNet block in \cite{8322184}. 
Given that the quantization operation is nondifferentiable, its gradient can be set to be one as in \cite{guo2020dl,guoJSAC}.
The entire feedback process can be formulated as
\begin{equation}
\label{totalProcess}
\hat{{\bf h}}_{\rm a} =  f_{\rm de}\Big(\big(\mathcal{Q}(f_{\rm en}({\bf h}_{\rm a},\Theta_{\rm en})\big),\Theta_{\rm de}\Big) ,
\end{equation}
where $f_{\rm en}(\cdot)$ and $f_{\rm de}(\cdot)$ denote the compression and reconstruction operations at the UE and the BS, respectively; $\Theta_{\rm en}$ and $\Theta_{\rm de}$ represent the NN weights of the encoder and the decoder, respectively; and $\mathcal{Q}(\cdot)$ is the quantization operation.
The entire feedback process can be trained using an end-to-end approach, and the loss function is mean squared error (MSE).

\section{Downlink pilot design and channel estimation}
\label{s3}
\subsection{Joint pilot design and channel estimation by using NNs}
\begin{figure*}[t]
    \centering 
    \includegraphics[scale=0.55]{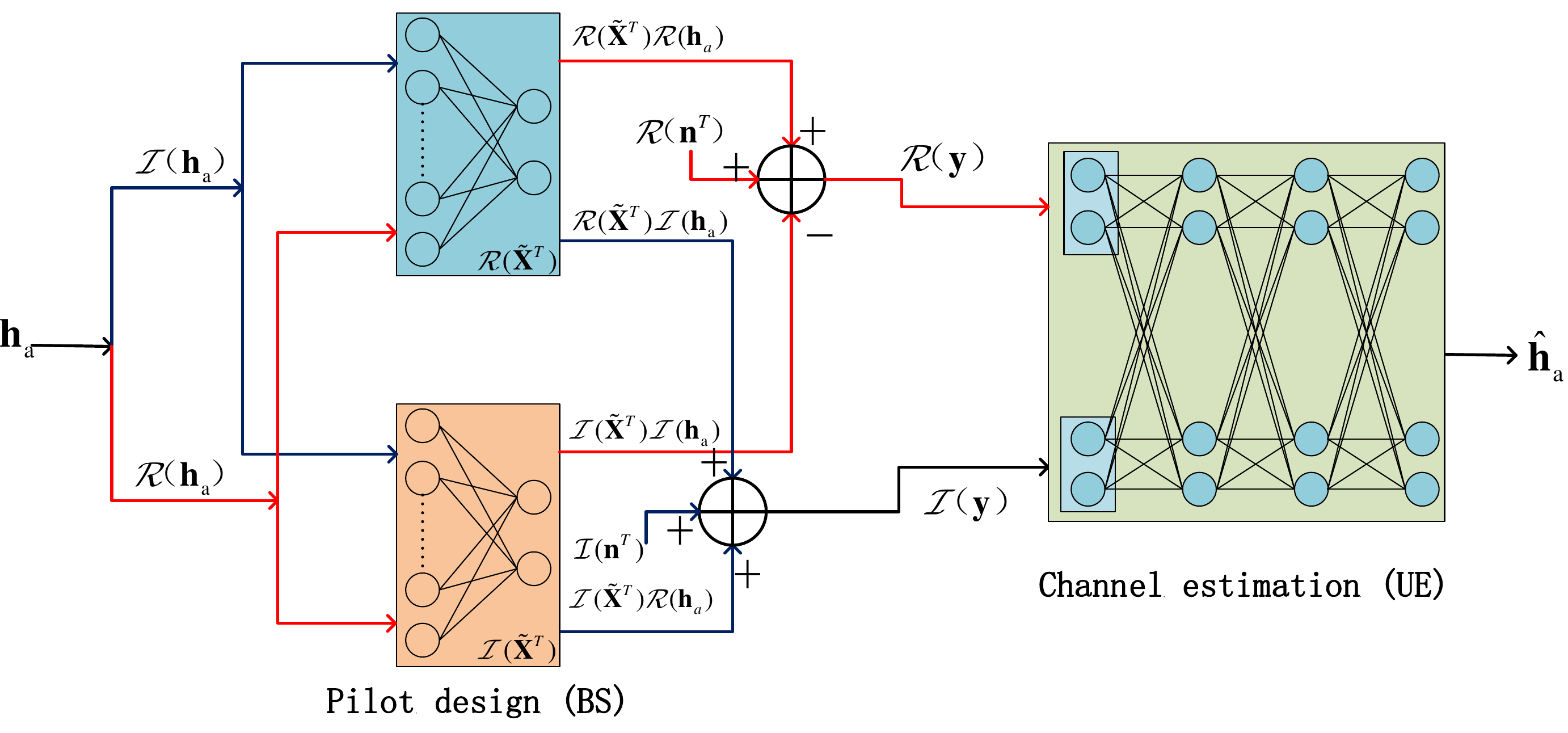}
    \caption{\label{Cpilot}Illustration of the DL-based joint pilot design and channel estimation framework. With reference to \cite{8861085}, two FC layers at the BS represent the real and imaginary parts of the pilot signals.}  
\end{figure*}
From the perspective of CS theory, (\ref{transmission}) can be regarded as a standard CS problem, where $\widetilde{X}$ and $ {\bf h}_{\rm a}$ are the sampling matrix and the sparse signal, respectively.
In image processing, a single NN layer, such as FC and convolutional layers, can play the role of a sampling matrix.
For example, the convolutional neural layer in \cite{8765626} replaces the sensing matrix and realizes compression.
The kernel size and  the filter number of the convolutional layer are the same as those of the input image block and output dimension, respectively.
However, these types of methods cannot be directly applied to communication problems because most matrices in communications are in the form of complex numbers.
To tackle this problem, \cite{8861085} uses two FC layers to realize compression of a complex matrix for device activity detection.
The simulation results in \cite{8861085,9174792} show that the learned sampling matrix can greatly improve recovery performance via learning the environment from the training channel data.
Inspired by this work, we propose a DL-based joint pilot design and channel estimation framework, as shown in Fig. \ref{Cpilot}.
We name this NN framework PEnet (\underline{P}ilot design and channel \underline{E}stimation NN framework).

The equation for the complex number in (\ref{transmission}) can be expressed equivalently through the following two equations for the real one as
\begin{equation}
\mathcal{R}({\bf y})  =\big ( (\mathcal{R}({\widetilde{X}^T }) \mathcal{R}({\bf h}_{\rm a}) -\mathcal{I}({\widetilde{X}^T }) \mathcal{I}({\bf h}_{\rm a}) ,
+\mathcal{R}({\bf n}^T)   \big )^T,
\end{equation}
\begin{equation}
\mathcal{I}({\bf y})  =\big ( (\mathcal{I}({\widetilde{X}^T }) \mathcal{R}({\bf h}_{\rm a}) +\mathcal{R}({\widetilde{X}^T }) \mathcal{I}({\bf h}_{\rm a}) 
+\mathcal{I}({\bf n}^T)   \big )^T,
\end{equation}
where $\mathcal{R}(\cdot)$ and $\mathcal{I}(\cdot)$ represent the real and imaginary parts of a complex number, respectively.
As shown in Fig. \ref{Cpilot}, on the basis of the two equations, the pilot design can be realized using two FC layers, which mimic the linear relations with real coefficient matrices, i.e.,  $\mathcal{R}({\widetilde{X}^T })$ and $\mathcal{I}({\widetilde{X}^T })$.
Each FC layer has no basis of nonlinear activation functions, and its input and output neuron numbers are $N_{\rm t}$ and $M$, respectively.
During the test phase, the weight of the connection between the $i$-th input neuron and $j$-th output neuron denotes the $(i,j)$-th element in the corresponding coefficient matrices, i.e.,  $\mathcal{R}({\widetilde{X}^T })$ or $\mathcal{I}({\widetilde{X}^T })$.
Unlike the conventional NNs, the two FC layers for the pilot design are executed twice during the training and test phases.
Therefore, we set the weights in the two FC layers as reusable in TensorFlow.
Channel estimation is realized using NNs, which consist of several FC layers, whose inputs are the concatenation of the real and imaginary parts of the received signal $\bf y$ at the UE and outputs are the real and imaginary parts of the estimated channel.
Additional details about the NNs for channel estimation are provided in Section \ref{pilotFramework001}.

\subsection{Uplink-aided downlink pilot design and channel estimation}
\subsubsection{Motivation}
In FDD massive MIMO systems, large training period $M$ can lead to enhanced channel estimation at the UE at the expense of substantial downlink resources, thereby leaving few time slots for actual data transmission.
Therefore, a pilot design strategy should be designed to alleviate this overhead.
In TDD massive MIMO systems, the downlink CSI can be inferred from the uplink one by exploiting the reciprocity among bidirectional channels. 
As mentioned in Section \ref{2A}, a correlation exists between the downlink and uplink channels in the angular domain if the duplex distance is not large.
The bidirectional channels share a common spatial structure in the magnitude of the angular domain.
Inspired by this condition, we can introduce the uplink channel magnitude information into the downlink pilot design process.
From the perspective of CS, the reciprocity in the angular domain translates to the same locations of nonzero entries in the angle-domain channel \cite{8697125}, i.e., 
\begin{equation}
\rm supp({{\bf h}}_{\rm a} ^{\rm ul}) =supp({{\bf h}}_{\rm a} ^{\rm dl}),
\end{equation}
where ${{\bf h}}_{\rm a} ^{\rm ul}$ and ${{\bf h}}_{\rm a} ^{\rm dl}$ are the channels in the angular domain of the uplink and the downlink, respectively; and $\rm supp(\cdot)$ represents the set of indices such that the corresponding entries are over $\epsilon$, which is slightly over zero.
Consequently, if we have obtained the uplink channel ${{\bf h}}_{\rm a} ^{\rm up}$ and the corresponding common support information, we can design dedicated pilot for the downlink channel estimation, which is based on the common support information.
\subsubsection{NN framework}
\label{pilotFramework001}
\begin{figure*}[t]
    \centering 
    \includegraphics[scale=0.75]{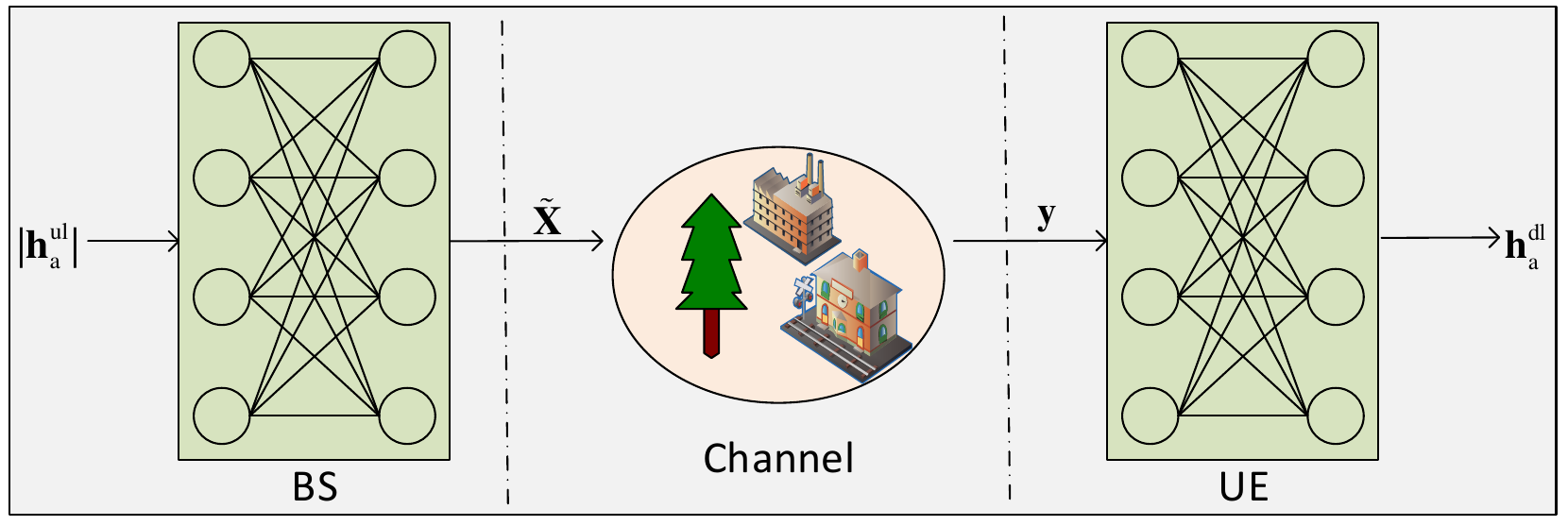}
    \caption{\label{pilotFramework}Illustration of the DL-based uplink-aided downlink pilot design and channel estimation framework, UpAid-PEnet.}  
\end{figure*}

Fig. \ref{pilotFramework} shows the framework for DL-based \underline{up}link-\underline{aid}ed downlink \underline{p}ilot design and channel \underline{e}stimation, called UpAid-PEnet .
Once the BS obtains the uplink CSI ${{\bf h}}_{\rm a} ^{\rm ul}$ through channel estimation, it generates downlink pilot signals $\widetilde{X}$ from the uplink channel magnitude, i.e., $| {{\bf h}}_{\rm a} ^{\rm ul} |$.
This step is realized using NNs, as in Fig. \ref{pilotFramework}.
The input of the NNs at the BS is the uplink channel magnitude $| {{\bf h}}_{\rm a} ^{\rm ul} |$, and the output is the pilot signal $\widetilde{X}$.
The BS transmits the pilot signals to the UE, which can be formulated as (\ref{transmission}).
Then, the UE receives the signal $\bf y$ and estimates the downlink channel from $\bf y$ also by using NNs.
The input of the NNs at the UE is the signal $\bf y$ and the output is the downlink channel ${\bf h}_{a}^{\rm dl}$.

The NNs consist of FC layers
\footnote{Since the goal of this paper is not designing novel NN architecture, we use simple FC layers.},
and the activation function used here is SELU.
Weight number and floating point operations (FLOPs) are the most used values to describe the NN complexity.
Here, we calculate the weight number and the number of FLOPs of the proposed NNs.
According \cite{guoJSAC}, the weight number and FLOPs of FC layers can be calculated as
\begin{equation}
\label{numberP}
N_{\rm FC} = N_{\rm o} \times (N_{\rm i}+1),
\end{equation}
\begin{equation}
\label{numberF}
{\rm FLOPs_{FC}} = N_{\rm o} \times (2N_{\rm i} -1),
\end{equation}
where $N_{\rm i} $ and $N_{\rm o} $ denote the input and output dimensions, respectively.
The detailed NN architecture and its complexity, including the parameter number and FLOPs, are provided in Table \ref{NN1}.
The input dimension of the channel estimation NNs is $2M\times 1$, which is decided by the pilot length. 
From the table, the total NN weight number of the proposed UpAid-PEnet is $(22+4M){N_{\rm t}^2} + (14+6M){N_{\rm t}}$ and the total FLOP number is $(44+8M){N_{\rm t}^2} + (6M-14){N_{\rm t}}$.

\begin{table*}[t]
\caption{\label{NN1}NN details of the DL-based uplink-aided downlink pilot design and channel estimation, UpAid-PEnet.}
\centering
\resizebox{\textwidth}{!}{
\begin{threeparttable}
\begin{tabular}{c|ccccc}
\hline \hline
                         & Layer name   & Output shape        & Activation operation *& Parameter number                        & FLOPs                                    \\ \hline \hline
\multirow{4}{*}{Pilot design (BS)} & Input        & $N_{\rm t}\times1$ & \textbackslash{} & 0                                       & 0                                        \\
                         & FC1          & $2N_{\rm t}\times1$ & SELU     & $2N_{\rm t}\times(N_{\rm t}+1)$        & $(2N_{\rm t}-1)\times 2N_{\rm t}$        \\
                         & FC2          & $2N_{\rm t}\times1$ &SELU      & $2N_{\rm t}\times(2N_{\rm t}+1)$        & $(4N_{\rm t}-1)\times 2N_{\rm t}$        \\
                         & FC3          & $2N_{\rm t}\times1$ &SELU      & $2N_{\rm t}\times(2N_{\rm t}+1)$        & $(4N_{\rm t}-1)\times 2N_{\rm t}$        \\
                         & FC4          & $2N_{\rm t} M \times  1$   & Tanh             & $2N_{\rm t} M\times(2N_{\rm t}+1)$ & $(4N_{\rm t}-1)\times 2N_{\rm t}M$ \\  \hline \hline
\multirow{4}{*}{Channel estimation (UE)} & FC5          & $2N_{\rm t}\times1$ & SELU & $2N_{\rm t}\times(2M+1)$   & $(4M-1)\times 2N_{\rm t}$   \\
                         & FC6          & $2N_{\rm t}\times1$ & SELU     & $2N_{\rm t}\times(2N_{\rm t}+1)$        & $(4N_{\rm t}-1)\times 2N_{\rm t}$        \\
                         & FC7          & $2N_{\rm t}\times1$  &SELU& $2N_{\rm t}\times(2N_{\rm t}+1)$         & $(4N_{\rm t}-1)\times2N_{\rm t}$         \\
                         &FC8         & $2N_{\rm t}\times1$  &Tanh& $2N_{\rm t}\times(2N_{\rm t}+1)$         & $(4N_{\rm t}-1)\times2N_{\rm t}$         \\ \hline \hline
\end{tabular}
\begin{tablenotes}
        \footnotesize
        \item[*] The FLOPs number of the activation operation is neglected since it is much smaller than that of the corresponding FC layer.
 \end{tablenotes}
\end{threeparttable}
}
\end{table*}

Although the pilot design and the channel estimation modules are separated during the inference phase, they are optimized together by using an end-to-end approach during the training phase, and the imperfect signal transmission process (\ref{transmission}) is considered.
Additional details about the NN training are provided in Section \ref{s6}.

\section{Uplink-aided downlink CSI feedback}
\label{s4}
In this section, we explain the motivation of the uplink-aided CSI feedback and introduce the NN framework, called UpAid-FBnet.
\subsection{Motivation}
\begin{figure*}[t]
    \centering 
    \includegraphics[scale=1.1]{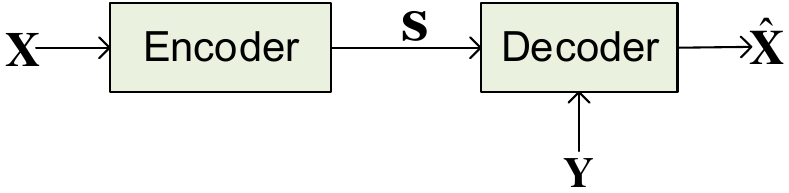}
    \caption{\label{dsc}Illustration of asymmetric DSC. The code word of $\bf X$ and the correlated source $\bf Y$ are both available to the decoder.}  
\end{figure*}

Asymmetric distributed source coding (DSC) is widely used in source-coding scenarios and a source $\bf X$ may be coded given a correlated source $\bf Y$, which is only available to the decoder, as shown in Fig. \ref{dsc}.
The framework in Fig. \ref{dsc} has been applied to many DL-based algorithms.
For example, in magnetic resonance imaging (MRI), multiple protocols exist during one exam, and MR images are highly correlated.
The authors in \cite{Zhou_2020_CVPR} propose an NN framework, called DuDoRNet, which embeds the prior of a fully sampled short protocol to undersampled MRI reconstruction with a long imaging protocol.
The CSI acquisition in FDD systems has a similar characteristic in which the uplink and downlink channels in the angular domain have some correlation if the duplex distance is not large.
The authors in \cite{8638509} observe the correlation in magnitude among bidirectional channels in the delay domain.
The magnitude and phase of the downlink CSI are fed back to the BS separately, thereby leading to a bit allocation problem that has not been solved.
To exploit the bidirectional correlation in angular domain and avoid the bit allocation problem, feeding back CSI magnitude and phase together and introducing the magnitude information of the uplink at the BS are reasonable.

\subsection{NN framework}

\begin{figure*}[t]
    \centering 
    \includegraphics[scale=1.2]{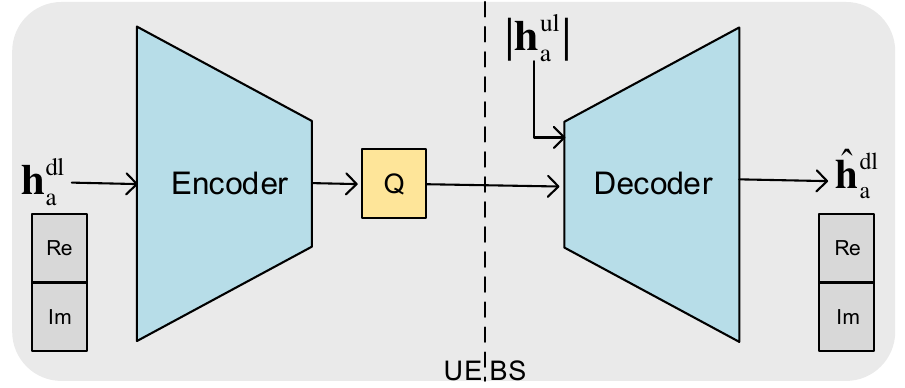}
    \caption{\label{FeedbackFramework}Illustration of the DL-based uplink-aided downlink CSI feedback framework, UpAid-FBnet.}  
\end{figure*}
In this part, we assume that the UE and the BS have obtained perfect bidirectional channels ${{\bf h}}_{\rm a} ^{\rm dl}$ and ${{\bf h}}_{\rm a} ^{\rm ul}$, respectively.
Fig. \ref{FeedbackFramework} illustrates the DL-based \underline{up}link-\underline{aid}ed downlink CSI \underline{f}eed\underline{b}ack framework, called  UpAid-FBnet, which is a specific case of the asymmetric DSC in Fig. \ref{dsc}.
From this figure, unlike the frameworks in \cite{guo2020dl,8638509}, which separately feed back the phase and the magnitude of the channel to exploit the correlation in the magnitude domain, the input of the NNs in the proposed framework is the complex downlink channel ${{\bf h}}_{\rm a} ^{\rm dl}$, whose real and imaginary parts are concatenated. 
At the UE, the encoder compresses the complex downlink channel ${{\bf h}}_{\rm a} ^{\rm dl}$ by using NNs and then quantizes the code words with 32-bit floating point numbers by using a uniform quantizer. Considering that the quantization operation is nondifferentiable, we set its gradient to be one during the training phase, as in \cite{guo2020dl,guoJSAC}.
Once the BS obtains the feedback, it concatenates the quantized measurement vector and the uplink channel magnitude $|{{\bf h}}_{\rm a} ^{\rm ul}|$ and sends them to the decoder.
The output of the decoder is the reconstructed complex downlink CSI ${{\bf h}}_{\rm a} ^{\rm dl}$.

Table \ref{NN2} shows the NN details of the uplink-aided feedback framework, which also consists of FC layers followed by SELU and tanh activation function.
Different from the encoder for pilot design, the output of the encoder in this task should be quantized to generate bitstreams for uplink transmission.
Following the setting in\cite{guo2020dl}, the quantization bit number is set to four.
Therefore, when the feedback bit number is $N_{\rm bits}$, the downlink CSI is first compressed $4\times 2N_{\rm t}  / N_{\rm bits}$ times.
For example, when the antenna number at the BS is 64, downlink CSI has 128 elements and the feedback bit number is 32, the CSI should be compressed $4 \times 128 /32 =16$ times, and the output of the decoder is an $8\times 1$ vector. Then, each element in this vector is quantized by four bits, accounting for 32-bit feedback overhead.
The parameter number of the NNs and the number of FLOPs are $16N_{\rm t}^2 + (\frac{3}{4}N_{\rm bits} + 9)N_{\rm t} + \frac{1}{4} N_{\rm bits}$ and $32N_{\rm t}^2 + (\frac{3}{2}N_{\rm bits} - 9)N_{\rm t} - \frac{1}{4} N_{\rm bits}$, respectively.
\begin{table*}[t]
\caption{\label{NN2}NN details of the DL-based uplink-aided downlink CSI feedback, UpAid-FBnet.}
\centering
\resizebox{\textwidth}{!}{
\begin{threeparttable}
\begin{tabular}{c|ccccc}
\hline \hline
                         & Layer name   & Output shape        & Activation operation & Parameter number                        & FLOPs                                    \\ \hline \hline
\multirow{4}{*}{Encoder (UE)} & Input1      & $2N_{\rm t}\times1$ & \textbackslash{} & 0                                       & 0                                        \\
                         & FC1          & $N_{\rm t}\times1$ & SELU     & $N_{\rm t}\times(2N_{\rm t}+1)$        & $(4N_{\rm t}-1)\times N_{\rm t}$        \\
                         & FC2          & $N_{\rm bits}/4\times1$ &Tanh      & $N_{\rm bits}/4\times(N_{\rm t}+1)$        & $(2N_{\rm t}-1)\times N_{\rm bits}/4$          \\ \hline \hline
\multirow{4}{*}{Decoder (BS)} & Input2    & $N_{\rm t}\times1$ & \textbackslash{} & 0                                       & 0                                        \\
& Concat *  & $(N_{\rm t}+N_{\rm bits}/4)\times1$ & \textbackslash{} & 0                                       & 0                                        \\
              &FC3& $2N_{\rm t}\times1$ & SELU & $2N_{\rm t}\times(N_{\rm t}+ N_{\rm bits}/4+1)$   & $(2N_{\rm t}+ N_{\rm bits}/2-1)\times 2N_{\rm t}$   \\
                         & FC4          & $2N_{\rm t}\times1$ & SELU     & $2N_{\rm t}\times(2N_{\rm t}+1)$        & $(4N_{\rm t}-1)\times 2N_{\rm t}$        \\
                         & FC5         & $2N_{\rm t}\times1$  &SELU& $2N_{\rm t}\times(2N_{\rm t}+1)$         & $(4N_{\rm t}-1)\times2N_{\rm t}$         \\
                         &FC6       & $2N_{\rm t}\times1$  &Tanh& $2N_{\rm t}\times(2N_{\rm t}+1)$         & $(4N_{\rm t}-1)\times2N_{\rm t}$         \\ \hline \hline
\end{tabular}
\begin{tablenotes}
        \footnotesize
        \item[*] The quantization operation and the FLOPs of the concat layer are neglected in this table.
 \end{tablenotes}
\end{threeparttable}
}
\end{table*}

\section{Uplink-aided downlink CSI acquisition}
\label{s5}
In this section, after explaining the motivation for uplink-aided downlink CSI acquisition, we introduce two types of CSI acquisition frameworks and provide their NN architecture and complexity. Given the high NN complexity, NN weight pruning is applied to the proposed NNs.

\subsection{Motivation}
In most downlink CSI acquisition frameworks, including conventional algorithms and DL-based methods, the UE first estimates the downlink channel and then feeds it back to the BS.
That is, the feedback and joint pilot transmission and channel estimation are separately designed.
In these works, no extra gains can be obtained if the entire CSI acquisition framework is jointly designed and optimized.
For example, in \cite{chenTVT}, CEFnet, which estimates and feeds back the downlink channel at the UE subsequently, outperforms PFnet, which directly feeds back the received pilot signal to the BS.
However, some papers suggest directly quantizing and feeding back the pilot signal \cite{8954618}.
In these works, extra expert knowledge, such as the correlation among nearby UE, is introduced to estimate the downlink channel at the BS.
For instance, in \cite{8954618}, the BS transmits pilots in the downlink, which are subsequently quantized at each UE and fed back to the BS via an uplink channel.
Then, the BS utilizes these feedback pilots to estimate each UE's  downlink channel by exploiting the channel’s hidden joint sparsity for channel estimation by using distributed CS algorithms.
Compared with the former CSI acquisition framework, this method can further reduce the feedback overhead.

Similar to that work \cite{8954618}, in our considered scenario, additional side information, i.e., the correlation among the bidirectional channels, is introduced into the CSI acquisition.
Therefore, different from that in \cite{chenTVT}, additional performance gains may be achieved if the channel estimation and the feedback are jointly designed and trained as an entire task by using an end-to-end approach.

\subsection{NN framework}
\label{CAnetNN}
In this subsection, we introduce two types of uplink-aided downlink CSI acquisition frameworks, CAnet-S and CAnet-J, which acquire the downlink CSI by using some separated modules and an end-to-end approach, respectively.

\subsubsection{Framework of CAnet-S}
\begin{figure*}[t]
    \centering 
    \includegraphics[scale=0.45]{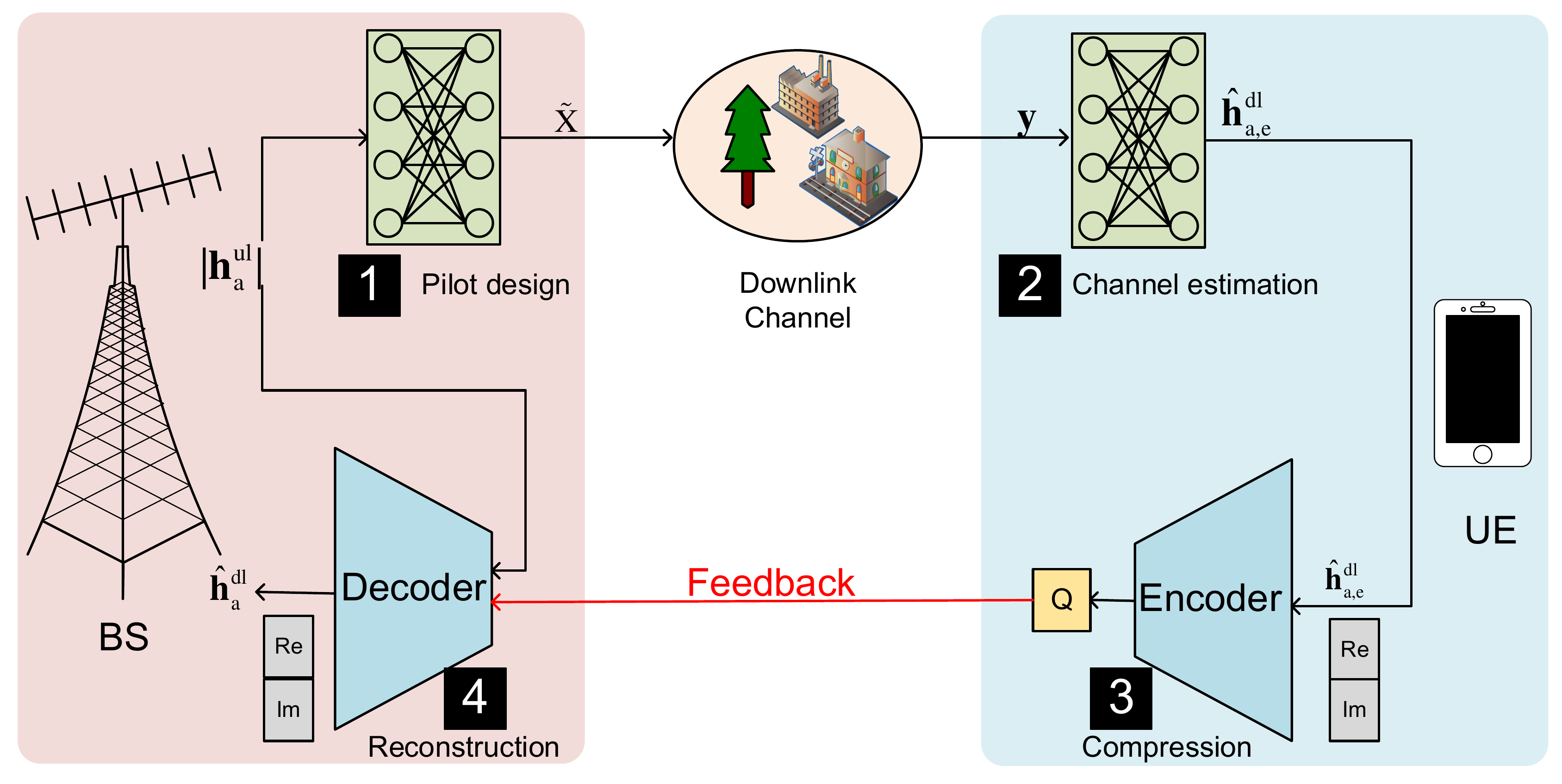}
    \caption{\label{CAnet-S}Flowchart of the DL-based uplink-aided downlink CSI acquisition framework, CAnet-S, which realizes CSI acquisition gradually.}  
\end{figure*}

CAnet-S is a combination of the above proposed UpAid-PEnet and UpAid-FBnet.
As shown in Fig. \ref{CAnet-S}, the BS can obtain the downlink CSI ${{\bf h}}_{\rm a} ^{\rm dl}$ via four main steps.
First, the BS designs the pilots in accordance with the uplink channel magnitude $| {{\bf h}}_{\rm a} ^{\rm ul} |$.
Then, the UE estimates the downlink channel.
Subsequently, the UE feeds back the compressed and quantized CSI to the BS.
Lastly, the BS reconstructs the downlink channel by using the feedback information and the  uplink channel magnitude information.
The received signal and the estimated downlink channel at the UE can be expressed as 
\begin{align}
{\bf y} &= ({\bf h}_{\rm a}^{\rm dl})^T{\rm f_{PD}}(| {{\bf h}}_{\rm a} ^{\rm ul} |,\Theta_{\rm PD}) + {\bf n},\\
{\hat {\bf h}}_{\rm a,e}^{\rm dl} &= {\rm f_{CE}} (   {\bf y} ,\Theta_{\rm CE} ),
\end{align}
where ${\rm f_{PD}}(\cdot)$ and $\Theta_{\rm PD}$ represent the pilot design operation and its NN parameter, respectively and ${\rm f_{CE}}(\cdot)$ and $\Theta_{\rm CE}$ represent the channel estimation operation and its NN parameter, respectively.
The feedback code word and the reconstructed downlink at the BS can be written as
\begin{align}
{\bf s}_{\rm 1} &= \mathcal{Q}\big(  f_{\rm en} ( {\hat {\bf h}}_{\rm a,e}^{\rm dl},\Theta_{\rm en} )\big),
\\
{\hat {\bf h}}_{\rm a}^{\rm dl} &=  f_{\rm de} ( {\bf s}_{\rm 1} ,\Theta_{\rm de} ).
\end{align}
The training process is divided into two phases.
The optimization goal of the first phase is to minimize the channel estimation error as 
\begin{equation}
 \mathop{\rm minimize}_{\Theta_{\rm PD},\Theta_{\rm CE}} \ \ \Big \|  {{\bf h}}_{\rm a} ^{\rm dl}-   
 {\hat {\bf h}}_{\rm a,e}^{\rm dl}  \Big\|_2^2.
\end{equation}
The second one minimizes the feedback error, which is formulated as
\begin{equation}
 \mathop{\rm minimize}_{\Theta_{\rm en},\Theta_{\rm de}} \ \ \Big \|  {{\bf h}}_{\rm a} ^{\rm dl}-   
 {\hat {\bf h}}_{\rm a}^{\rm dl}  \Big\|_2^2.
\end{equation}

The complexity of the proposed CAnet-S, including the NN parameter number and the FLOP number, is the complexity sum of UpAid-PEnet and UpAid-FBnet.
Therefore, the parameter number is 
$(38+4M){N_{\rm t}^2} + (\frac{3}{4}N_{\rm bits} + 25+6M){N_{\rm t}}+ \frac{1}{4} N_{\rm bits}$ and the total FLOP number is $(76+8M){N_{\rm t}^2} - (23-6M-\frac{3}{2}N_{\rm bits} ){N_{\rm t}}- \frac{1}{4} N_{\rm bits}$.

\subsubsection{Framework of CAnet-J}

\begin{figure*}[t]
    \centering 
    \includegraphics[scale=0.45]{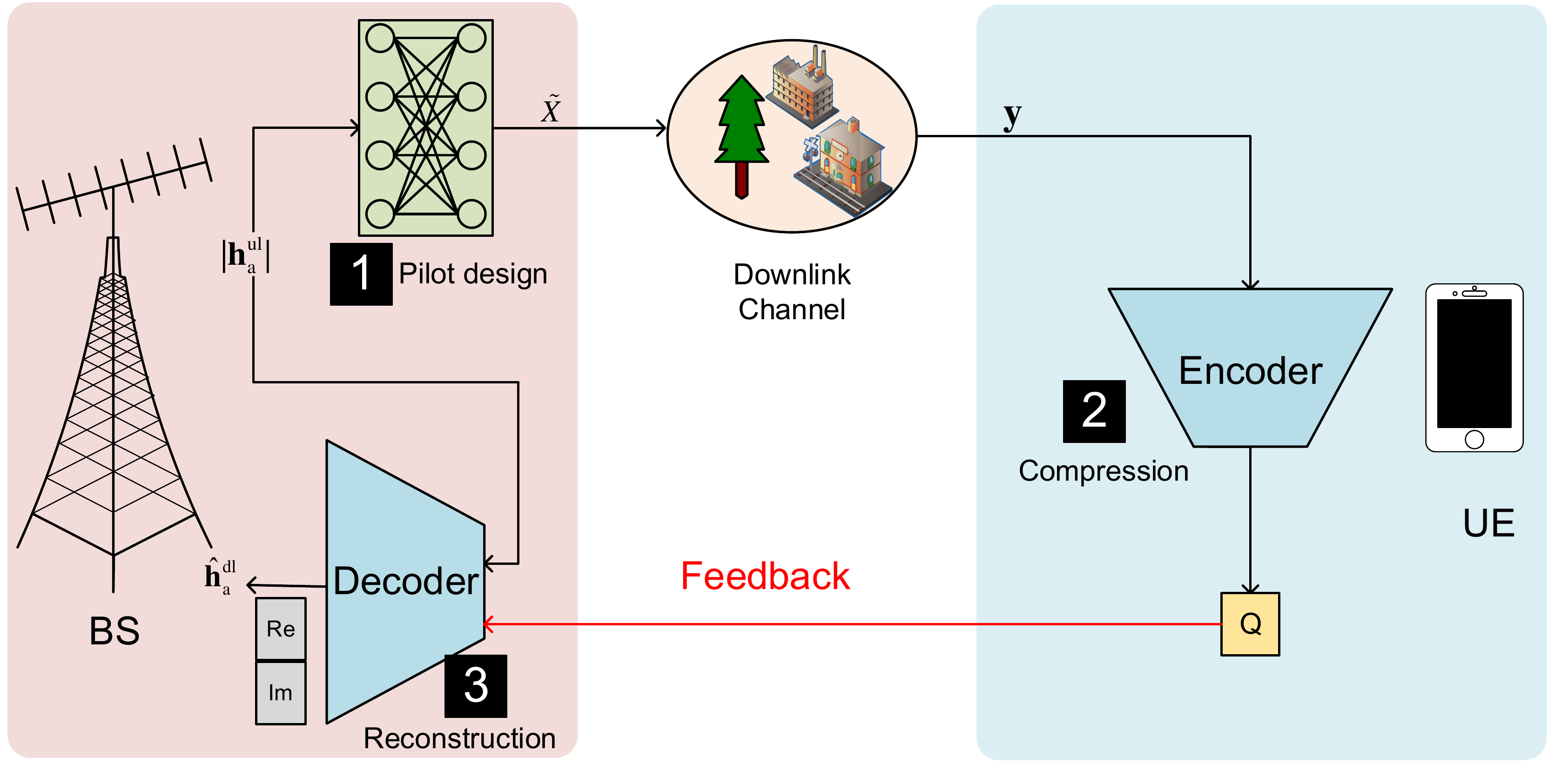}
    \caption{\label{CAnet-J}Flowchart of the DL-based uplink-aided downlink CSI acquisition framework, CAnet-J, which is optimized using an end-to-end approach.}  
\end{figure*}

Fig. \ref{CAnet-J} illustrates the flowchart of the DL-based uplink-aided downlink CSI acquisition framework, CAnet-J, which is optimized using an end-to-end approach.
From the figure, three main steps are implemented for FDD downlink CSI acquisition in the proposed CAnet-J framework.
Similar to the the process in CAnet-S, the BS first designs the pilot by exploiting the uplink channel magnitude information.
However, once the UE receives the pilot signal $\bf y$, it does not estimate the downlink channel but directly compresses, quantizes, and feeds back the pilot signal.
The compression operation is realized using an encoder, which only consists of one FC layer, followed by tanh activation function.
Lastly, the BS reconstructs the downlink channel by using the feedback information and the uplink channel magnitude information.
Therefore, the feedback code word and the reconstructed channel at the BS should be redefined as
\begin{equation}
{\bf s}_{\rm 2}  =  \mathcal{Q}\big(  f_{\rm en} ( {\bf y},\Theta_{\rm en} )\big)
\end{equation}
\begin{equation}
{\hat {\bf h}}_{\rm a}^{\rm dl} =  f_{\rm de} ( {\bf s}_{\rm 2} ,\Theta_{\rm de} ).
\end{equation}
The entire framework is trained using an end-to-end approach. The loss function is MSE and the optimization process can be formulated as
\begin{equation}
 \mathop{\rm minimize}_{\Theta_{\rm PD},\Theta_{\rm en},\Theta_{\rm de}} \ \ \Big \|  {{\bf h}}_{\rm a} ^{\rm dl}-   
 {\hat {\bf h}}_{\rm a}^{\rm dl}  \Big\|_2^2.
\end{equation}

Table \ref{NN33} provides the NN architecture and the complexity of the proposed CAnet-J.
The complexities of the proposed CAnet-J, including the NN parameter number and the FLOP number, are 
$(24+4M){N_{\rm t}^2} + (\frac{1}{2}N_{\rm bits} + 14+2M){N_{\rm t}}+ \frac{1}{2} MN_{\rm bits} + \frac{1}{4}  N_{\rm bits}$ and $(48+8M){N_{\rm t}^2} - (14+2M-N_{\rm bits} ){N_{\rm t}}+MN_{\rm bits}- \frac{1}{4} N_{\rm bits}$, respectively.
Compared with CAnet-S, CAnet-J does not need channel estimation and has a simpler encoder at the UE.
As a result, its complexity is much lower than that of CAnet-S, especially the NNs at the UE.

\begin{table*}[t]
\caption{\label{NN33}NN details of the DL-based uplink-aided downlink CSI acquisition framework, CAnet-J, which is optimized by an end-to-end approach.}
\centering
\resizebox{\textwidth}{!}{
\begin{threeparttable}
\begin{tabular}{c|ccccc}
\hline \hline
                         & Layer name   & Output shape        & Activation operation *& Parameter number                        & FLOPs                                    \\ \hline \hline
\multirow{5}{*}{Pilot design (BS)} & Input        & $N_{\rm t}\times1$ & \textbackslash{} & 0                                       & 0                                        \\
                         & FC1          & $2N_{\rm t}\times1$ & SELU     & $2N_{\rm t}\times(N_{\rm t}+1)$        & $(2N_{\rm t}-1)\times 2N_{\rm t}$        \\
                         & FC2          & $2N_{\rm t}\times1$ &SELU      & $2N_{\rm t}\times(2N_{\rm t}+1)$        & $(4N_{\rm t}-1)\times 2N_{\rm t}$        \\
                         & FC3          & $2N_{\rm t}\times1$ &SELU      & $2N_{\rm t}\times(2N_{\rm t}+1)$        & $(4N_{\rm t}-1)\times 2N_{\rm t}$        \\
                         & FC4          & $2N_{\rm t} M \times  1$   & Tanh             & $2N_{\rm t} M\times(2N_{\rm t}+1)$ & $(4N_{\rm t}-1)\times 2N_{\rm t}M$ \\  \hline \hline
\multirow{1}{*}{Encoder (UE)}            & FC5          & $N_{\rm bits}/4\times1$ &Tanh      & $N_{\rm bits}/4\times(2M+1)$        & $(4M-1)\times N_{\rm bits}/4$                 \\ \hline \hline
\multirow{4}{*}{Decoder (BS)} & Input2    & $N_{\rm t}\times1$ & \textbackslash{} & 0                                       & 0                                        \\
& Concat   & $(N_{\rm t}+N_{\rm bits}/4)\times1$ & \textbackslash{} & 0                                       & 0                                        \\
              &FC3& $2N_{\rm t}\times1$ & SELU & $2N_{\rm t}\times(N_{\rm t}+ N_{\rm bits}/4+1)$   & $(2N_{\rm t}+ N_{\rm bits}/2-1)\times 2N_{\rm t}$   \\
                         & FC4          & $2N_{\rm t}\times1$ & SELU     & $2N_{\rm t}\times(2N_{\rm t}+1)$        & $(4N_{\rm t}-1)\times 2N_{\rm t}$        \\
                         & FC5         & $2N_{\rm t}\times1$  &SELU& $2N_{\rm t}\times(2N_{\rm t}+1)$         & $(4N_{\rm t}-1)\times2N_{\rm t}$         \\
                         &FC6       & $2N_{\rm t}\times1$  &Tanh& $2N_{\rm t}\times(2N_{\rm t}+1)$         & $(4N_{\rm t}-1)\times2N_{\rm t}$         \\ \hline \hline
\end{tabular}
\end{threeparttable}
}
\end{table*}

\subsubsection{NN weight pruning}
Most existing works focus only on exploiting NNs to improve communication system performance but ignore the high complexity of NN-based communication algorithms, which hinders the deployment of DL-based algorithms in communications \cite{9136588}.
For example, when the antenna number at the BS, pilot length, and feedback bit number are 64, 16, and 32, respectively, in accordance with Table \ref{NN33}, the parameter number and the FLOPs of CAnet-J are 364,936 and 721,016, respectively, which require large model storage space and computational power.
NN compression, including NN quantization, weight pruning, and knowledge distillation, has been regarded as a potential technology to tackle the complexity challenge.
NN weight pruning and quantization techniques are widely used in FC layers, which are the main NN layers adopted in this work.
According to \cite{9136588,8924932}, no performance drop will occur if the quantization bit number of the NN weight is more than 4 or 5.
Therefore, we only apply the NN pruning technique to the proposed CAnet in this paper.

NN weight-pruning methods consist of two main types.
The first one prunes the NNs during the training, whereas the other one prunes and fine-tunes the NNs after pretraining.
In this work, we apply magnitude-based NN weight pruning to the NNs after pretraining by using the TensorFlow Model Optimization Toolkit \cite{tfmt}. 
In DL, sparse NN models are easy to compress, and zeros can be skipped during the inference, thereby improving latency for the equipment with restrictions on memory, processing, power consumption, model storage space, and network usage \cite{9136588}.
Compared with a method that prunes weights with small absolute values at once, such as in \cite{9136588}, NN weights in this work are gradually zeroed out to achieve NN sparsity during the training phase.

Specifically, a binary mask variable is added to each layer, which needs to be pruned.
These masks have the same shapes as the corresponding weight tensors and determine which NN weights can participate in the forward execution of the graph in the TensorFlow.
In each pruning step, we sort the weights by using their absolute values in the layer and mask to zero the smallest magnitude weights until the desired sparsity level $s$ is reached.
The sparsity level denotes the percentage of the pruned NN weights from the total NN weights.
During the backpropagation, the gradients flow through the binary masks and do not update the NN weights, which are masked in the forward execution. 
As mentioned before, the pruning is realized gradually and the sparsity level after each pruning operation should be designed.
In this work, we  adopt the automated gradual pruning mechanism proposed by \cite{zhu2017prune}.
In that mechanism, the sparsity level increases from an initial sparsity value $s_{\rm i}$ (usually zero) to a final predefined sparsity value $s_{\rm f}$ via $n$ pruning steps.
The adaptive sparsity is defined as
\begin{equation}
\label{eqP}
s_{\rm t} = s_{\rm f} + (s_{\rm i} - s_{\rm f})\Big(  1- \frac{t-t_0}{n \Delta t} \Big)^3 \qquad
\text { for}\quad  t  \in \{ t_0,t_0+\Delta t, \cdots,  t_0+n\Delta t   \},
\end{equation}
where $t_0$ and $\Delta t$ denote the start straining step and pruning frequency, respectively.
The intuition behind the sparsity update function (\ref{eqP}) is that the NNs are redundant at the beginning; hence, we prune the networks rapidly in the initial phase.
Additional details about this mechanism can be referred to \cite{zhu2017prune} .

\section{Simulation results and discussions}
\label{s6}
In this section, we first provide the channel generation and NN training details.
Then, we evaluate and discuss the proposed uplink-aided joint pilot design and channel estimation, channel feedback, and lightweight CSI acquisition frameworks.

\subsection{Simulation setting}

\subsubsection{Channel generation}

In this paper, we adopt the 3GPP SCM channel model \cite{3gpp996} to generate the bidirectional channels in an urban microcell.
Following the setting in \cite{8638509}, we set the uplink frequency to 5.1 GHz and the downlink frequency to 5.3 GHz.  
We assume that $N_{\rm c} = 3$ random scattering clusters range from $-\pi/2$ to $\pi /2$ and that each of the clusters contains $N_{\rm s}=20$ suppaths.
The BS is equipped with 64 ULA antennas, and the UE has a single-receiver antenna.
According to \cite{8697125,8383706}, we fix the frequency-independent parameter, such as AoD, and vary the complex gain of each path between the downlink and uplink channels.
We generate 1,000,000 CSI samples by using MATLAB, and the generated channel data is then divided into training, validation, and test datasets, with 80\%, 10\%, and 10\% channel samples, respectively.

\subsubsection{NN training}
NN training is implemented on Keras, a high-level library for TensorFlow v2.3.0, by using an NVIDIA DGX-2 workstation.
The batch size and the learning rate are set to 512 and 0.001, respectively.
We adopt the adaptive moment estimation (Adam) algorithm as an NN optimizer, whose parameter setting follows the default one \footnote{\url{https://tensorflow.google.cn/api_docs/python/tf/keras/optimizers/Adam}}.

\subsection{Performance of joint pilot design and channel estimation}

\begin{figure*}[t]
    \centering 
    \includegraphics[scale=0.700]{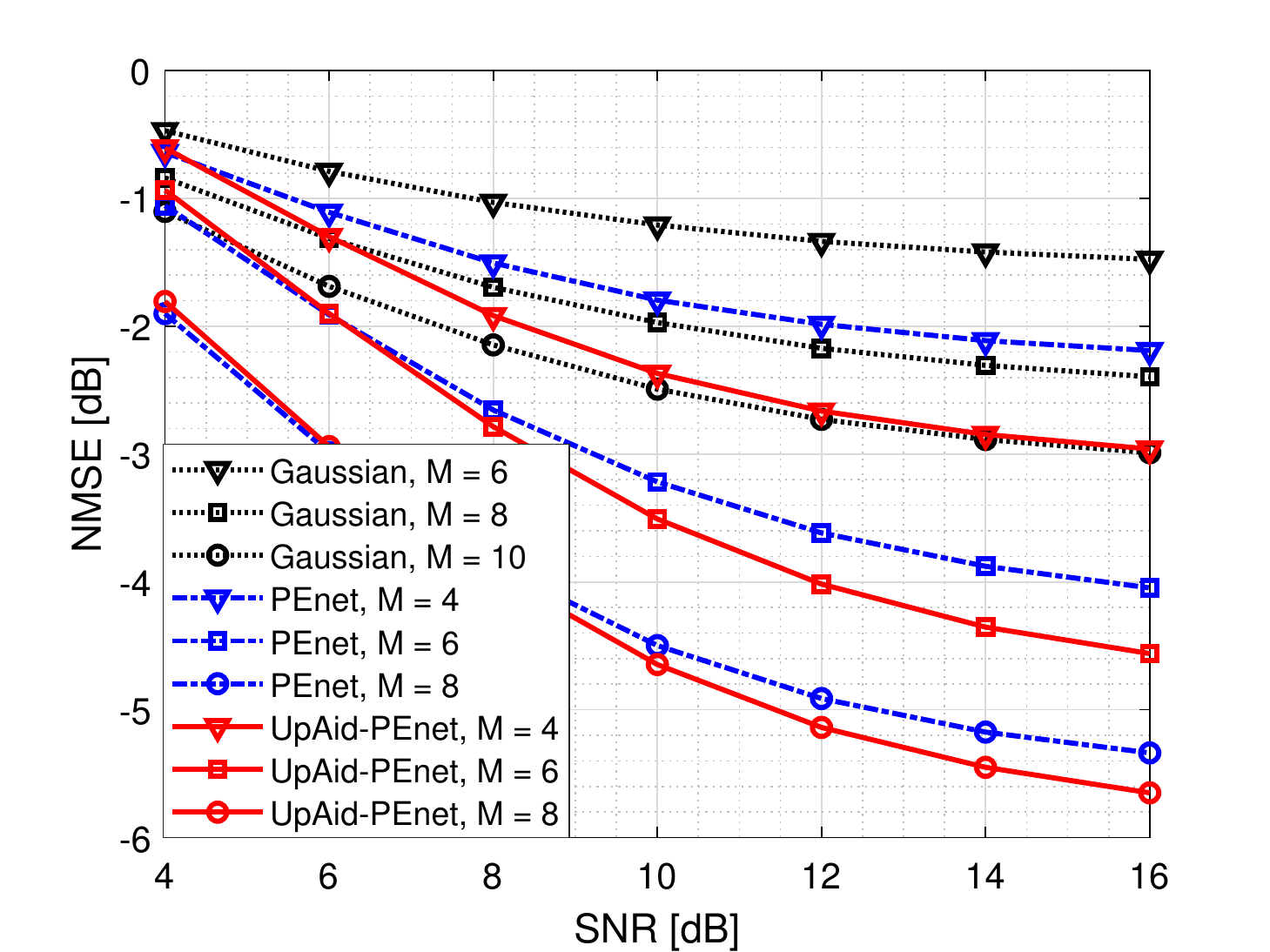}
    \caption{\label{UpPEnet-R} Comparison of NMSE performance versus SNRs between different pilot design and channel estimation schemes for massive MIMO systems. The training SNR is 10 dB.}  
\end{figure*}

The simulation results in \cite{8861085,9174792} show that the learned sampling matrix can greatly improve recovery performance compared with traditional algorithms, such as AMP and LASSO.
Therefore, we directly compare the proposed UpAid-PEnet with PEnet.
The channel estimation module of PEnet is the same as that of UpAid-PEnet, as indicated in Table \ref{NN1}.
Thus, the NNs at the UE for the two methods have the same complexity.
Fig. \ref{UpPEnet-R} compares the normalized MSE (NMSE) performance of the proposed UpAid-PEnet with that of PEnet against different SNRs.
In the figure, ``Gaussian'' represents the algorithm in which a random complex Gaussian matrix is regarded as the pilot signal $\widetilde{ X}$ and the channel estimation module is the same as those of PEnet and UpAid-PEnet.
The training SNR for them is 10 dB, and the test SNR varies from 4 dB to 16 dB.
Similar to the observation in \cite{8861085}, the methods with a learned measurement matrix, i.e.,  PEnet and UpAid-PEnet, outperform the method with a random Gaussian matrix by a large margin.
For example, when the test SNR is 12 dB, the UpAid-PEnet with 4 pilots has comparable NMSE performance with the ``Gaussian'' with 10 pilots, thereby reducing 60\% of the training overhead.
From the figure, the proposed UpAid-PEnet always performs better than the PEnet with the same pilot overhead.
With the decrease in pilot length $M$, the performance gap between the two proposed NNs widens.
When the pilot length $M$, i.e., downlink training, is sufficient, the necessity of the aid from the uplink channel will be weakened.
Meanwhile, the figure shows that the performance gain increases with the test SNR.
If the estimation SNR is very low, then the quality of the channel estimated at the UE will be low regardless of what pilots are adopted.

In UpAid-PEnet, the pilot depends on the uplink channel magnitude, and the UE has no knowledge about the uplink channel at all.
That is, the UE does not know the pilot signals at all,
which leads to a doubt on how the estimation module at the UE can work without the knowledge of the pilot signals.
We calculate the NMSE among the designed pilot signals.
The simulation results show that these pilot signals are different but similar, and the NMSE between two pilot signals is below -10 dB.
Therefore, the channel estimation module has known the main information about the transmitted pilot signal $\widetilde{X}$.
The NNs for pilot design may fine-tune the pilot signals in accordance with the downlink channel, thereby improving the performance in channel estimation.
From the perspective of CS, this gain is achieved using a well-designed adaptive measurement matrix, which is in the form of precoded pilot signals in this work.
\begin{figure*}[t]
    \centering 
    \includegraphics[scale=0.700]{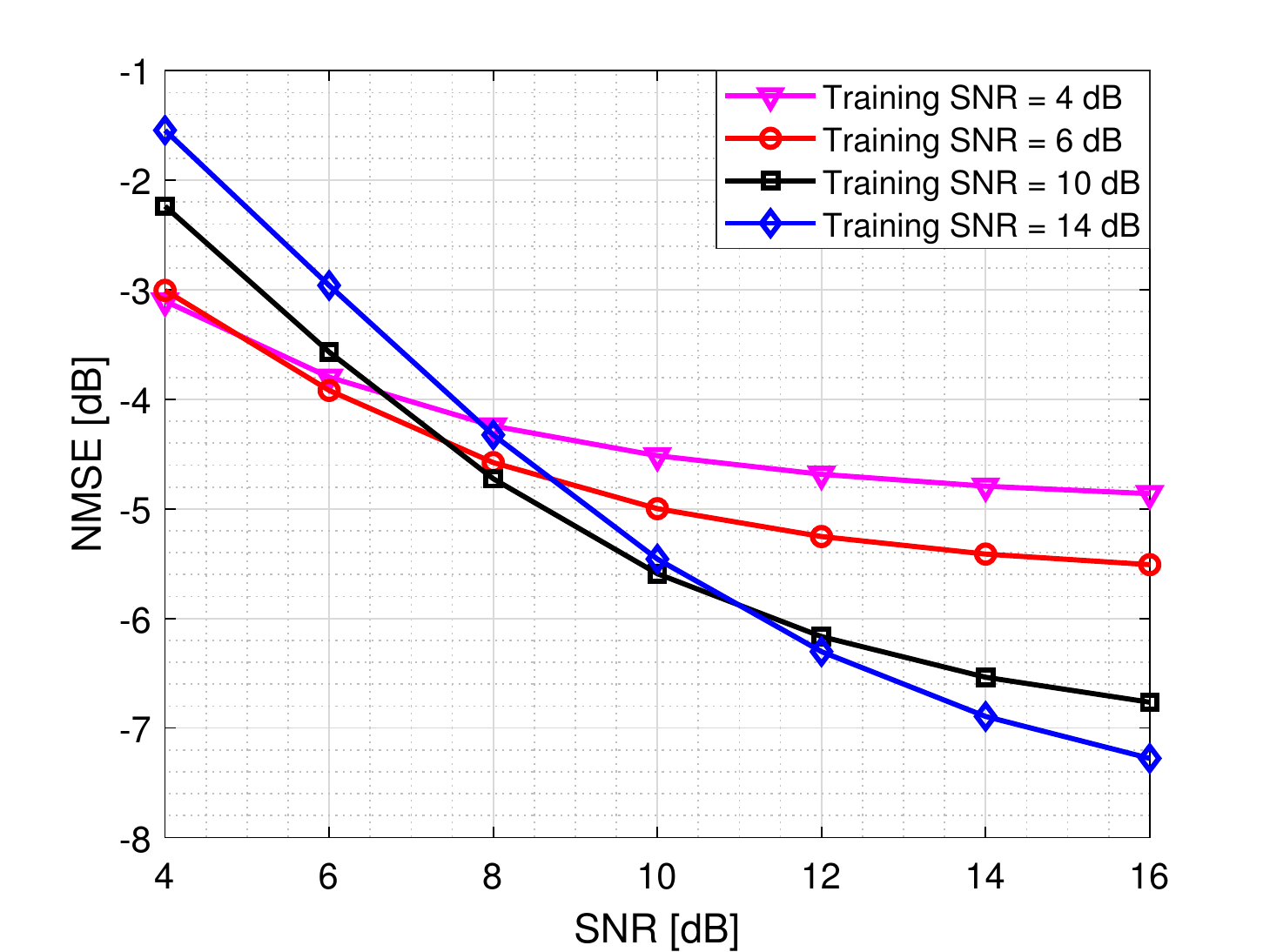}
    \caption{\label{UpPEnetRobust} NMSE performance versus SNRs of UpAid-PEnet under different training SNRs.}  
\end{figure*}

Fig. \ref{UpPEnetRobust} plots the robustness of the proposed UpAid-PEnet trained under different SNRs, i.e., 4, 6, 10, and 14 dB.
The simulation results show that when the training and test SNRs are close, the NNs perform efficiently.
To improve the NNs' robustness to SNR,  we can adopt the multi-SNR training scheme proposed in \cite{chenTVT} in which the training datasets are composed of the datasets for all the predefined SNR levels.
Nevertheless, considering that the goal of this study is not to improve the robustness, we do not apply multi-SNR training scheme to the proposed NNs.

\subsection{Performance of the uplink-aided downlink channel feedback: UpAid-FBnet}

\begin{figure*}[t]
    \centering 
    \includegraphics[scale=0.700]{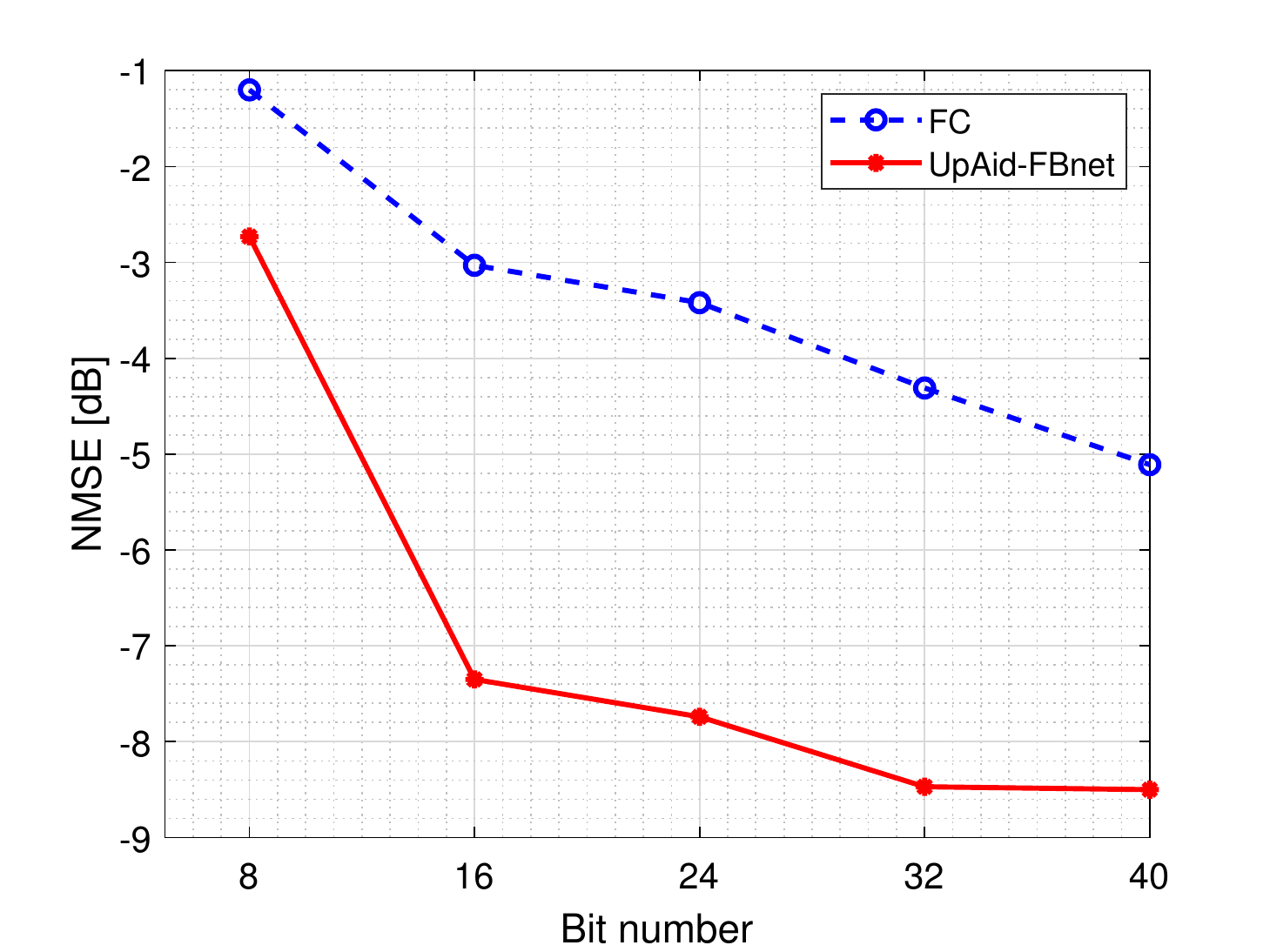}
    \caption{\label{UpFBnet-R} Comparison of NMSE performance versus feedback bit number among different feedback methods.}  
\end{figure*}

In this part, we assume that the UE and the BS have obtained the perfect downlink and uplink channels, respectively, and focus only on the feedback process.
The baseline algorithm, FC, has the same NN architecture as UpAid-FBnet but does not input the uplink information into the decoder at the BS.
Therefore, these two methods have nearly the same NN complexity.
Fig. \ref{UpFBnet-R} shows the NMSE performance comparison between FC and UpAid-FBnet with different feedback bit numbers.
Different from \cite{guo2020dl}, we evaluate the NMSE versus the total feedback bit number because the proposed strategy avoids the bit allocation problem.
As expected, the feedback accuracy increases with the feedback overhead.
The proposed uplink-aided feedback strategy outperforms the original one by a large margin, which verifies the necessity of exploiting uplink information during the feedback of the downlink channel. 

\subsection{Performance of uplink-aided downlink CSI acquisition: CAnet}
In this subsection, we compare the NMSE and the robustness performance of the two proposed CAnet frameworks.
Considering the high complexity of NN-based algorithms, we evaluate the effects of the NN compression technology on CAnet-J.

\subsubsection{NMSE performance of the proposed CAnet}
\label{ssD}

Given that the goal of this part is to compare the two proposed CAnet frameworks, the training and test SNRs are both 10 dB, and the pilot length $M$ is set to 6 and 8. 
Table \ref{CAnet-R} shows the NMSE comparison among AcqNet-J, CAnet-S, and  CAnet-J.
The framework of AcqNet-J is similar to that of CAnet-J and the pilot design is based on PEnet.
When the feedback bit number is low, the pilot length has a limited effect on the CSI quality.
For example, when the feedback bit number is 24, the NMSEs of CAnet-J with pilot length $M$=6, 8 are $-7.11$ and $-6.76$ dB, respectively, and those of CAnet-S are $-4.59$ and $-4.94$ dB, respectively.
The NMSE performance gap is relatively small.
However, the former one, which only sends $M=6$ pilots, reduces the training overhead by 25\%.
This result shows the inseparable relationship between the pilot length and the feedback bit number.
During the downlink channel acquisition, the two parts cannot be designed separately.
When the feedback overhead is extremely limited, the downlink CSI obtained at the BS must be of low quality even if the channel is well estimated at the UE at the expense of a high training overhead.
That is, if the feedback bit number is low, then the downlink channel need not be estimated well.

\begin{table*}[t]
\caption{\label{CAnet-R} NMSE performance versus feedback bit number comparison.}  
\centering
\begin{tabular}{c|c|cc}
\hline
$N_{\rm bits}$      & Methods     & $M=6$ & $M=8$ \\ \hline\hline
\multirow{3}{*}{16} & AcqNet-J & -3.82 &-3.95 \\
                    & CAnet-S & -3.88 & \textbf{-4.52} \\
                    & CAnet-J &\textbf{-4.36} & -4.32 \\ \hline
\multirow{3}{*}{20} &AcqNet-J & -5.28 & -5.01 \\
                    & CAnet-S                     & -4.37                     & -4.63                     \\
                    & CAnet-J                     & \textbf{-5.63 }                    & \textbf{-5.59}                     \\
                    \hline
\multirow{3}{*}{24} & ComAcqNet-J                     & -6.30                     & -6.45                     \\
                    & CAnet-S                     & -4.59                     & -4.94                     \\
                    & CAnet-J                     & \textbf{-7.11}                     & \textbf{-6.76}                    \\\hline
\end{tabular}
\end{table*}

As depicted in Fig. \ref{UpPEnet-R}, the NMSEs for the two pilot lengths are $-3.50$ and $-4.64$ dB when we consider only the channel estimation.
Nevertheless, from Table \ref{CAnet-R}, when the feedback bit number is over 16, the NMSEs of the CAnet-S are lower than them, i.e., the CSI estimated from the pilot signals at the UE is outperformed.
In other words, when the feedback bit number is over a threshold, the feedback process can even improve the downlink channel quality, which is due to the introduction of the uplink channel information into the decoder of the BS.

Then, we compare the two proposed CAnet frameworks with each other.
When the feedback bit number is over 16, CAnet-J performs much better than CAnet-S.
As described in Section \ref{CAnetNN}, from the perspective of NN architecture, CAnet-J and CAnet-S have similar architecture but the complexity of CAnet-S is slightly higher.
Both of them exploit the uplink channel information to design pilot and feedback modules.
The main difference between them is the training strategy; they train NNs via one step and two steps.
In the channel estimation phase of CAnet-S, the UE has no knowledge of the uplink channel.
Given that the UE has to estimate the full CSI, the received pilot signals have to contain all information about the downlink channel, which leads to the need for additional training and feedback overheads.
In CAnet-J, the UE does not need to estimate the full CSI and directly feeds back the received signals to the BS.
At the BS, the decoder reconstructs the downlink channel on the basis of the feedback and the uplink channel information.
In this framework, the UE does not need to feed back the information shared by bidirectional channels.
Given that all modules in CAnet-J are jointly trained, during the NN update process, i.e., backward propagation, the designed pilot is forced to focus on the channel information that is not shared by bidirectional channels, thereby reducing the training overhead and increasing the quality of the obtained downlink channel.

\subsubsection{Robustness performance of the proposed CAnet-J}

\begin{figure*}[t]
    \centering 
    \includegraphics[scale=0.700]{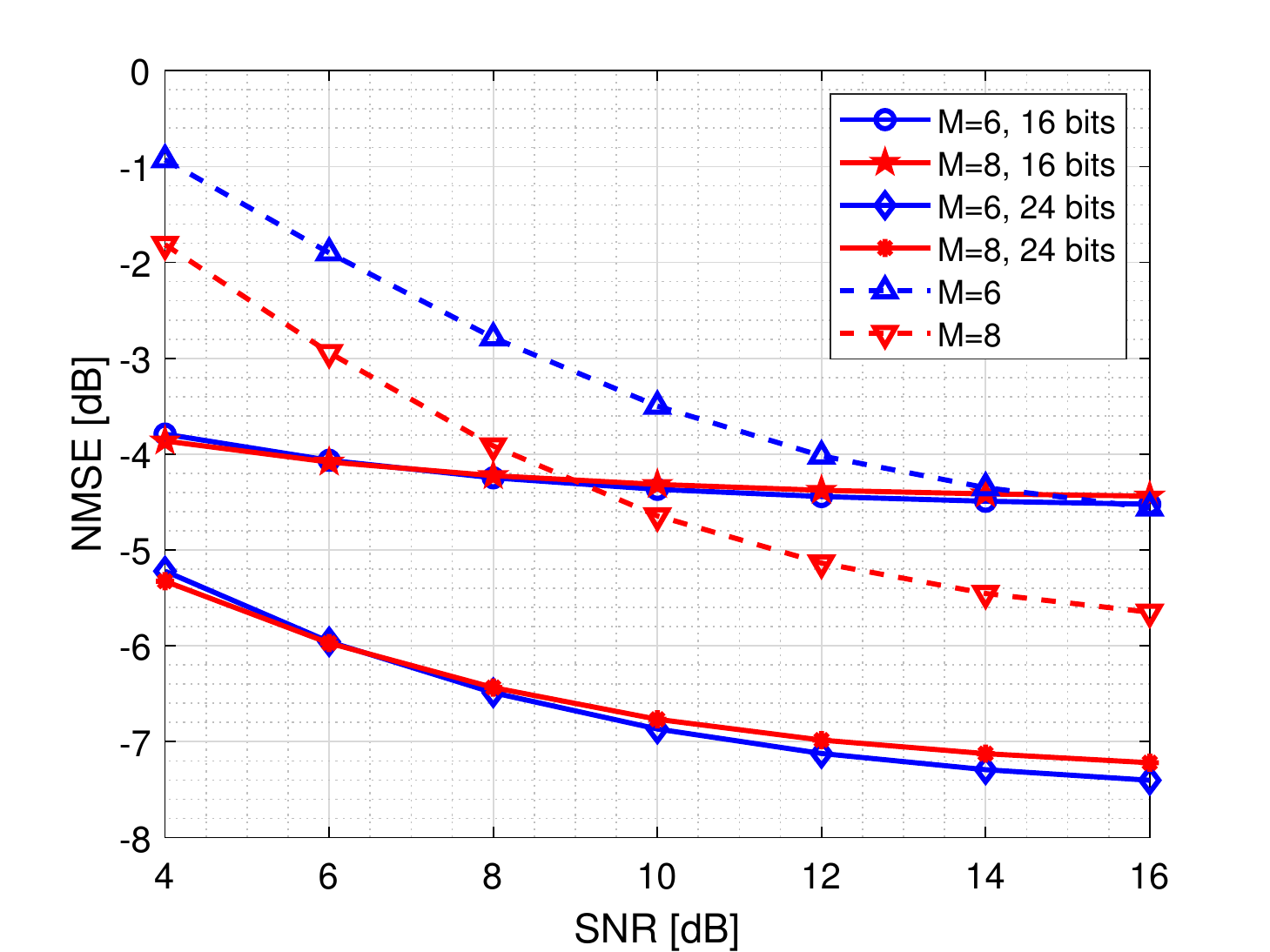}
    \caption{\label{CAnetRobust} NMSE performance versus SNR of the CAnet-J framework with different pilot lengths and feedback bit numbers. The training SNR is set as 10 dB. The dotted lines plot the performance of UpAid-PEnet, which does not consider the feedback process.}  
\end{figure*}

Fig. \ref{CAnetRobust} shows the robustness of the CAnet-J framework with different pilot lengths and feedback bit numbers. The training SNR is set to 10 dB.
The dotted line represents the performance of UpAid-PEnet, which does not consider the feedback process.
Compared with the channel estimation module, i.e., UpAid-PEnet, the proposed CAnet-J is more robust to channel noise.
For example, when the test SNR, the pilot length, and the feedback bit number are 6 dB, 8, and 24, respectively, the performance drops of UpAid-PEnet and CAnet-J are 1.70 and 0.80 dB, respectively.
When the SNR is very low, such as 4 dB and 6 dB, CAnet-J outperforms UpAid-PEnet by a large margin.
In this case, the received pilot signals are interrupted by high noise, and limited information can be exploited to estimate the downlink channel at the UE.
However, in the CAnet-J framework, the decoder at the BS can exploit the information not only from the compressed pilot signals but also from the high-quality uplink channel, which may help reconstruct a coarse downlink channel.
The simulation results in Section \ref{ssD} show that when the feedback bit number is 24, the pilot length need not be increased from 6 to 8 due to the limited feedback overhead.
Nonetheless, in consideration of the high channel noise during downlink training, as demonstrated in Fig. \ref{CAnetRobust}, the NNs with $M=8$ pilots perform better than those with $M=6$ pilots when the test SNR is 4 dB.
If the SNR is low, the pilot signals can contain insufficient information.
Consequently, considerable pilot signals are needed in this case.

\subsubsection{Pruned CAnet-J}

\begin{figure*}[t]
    \centering 
    \includegraphics[scale=0.700]{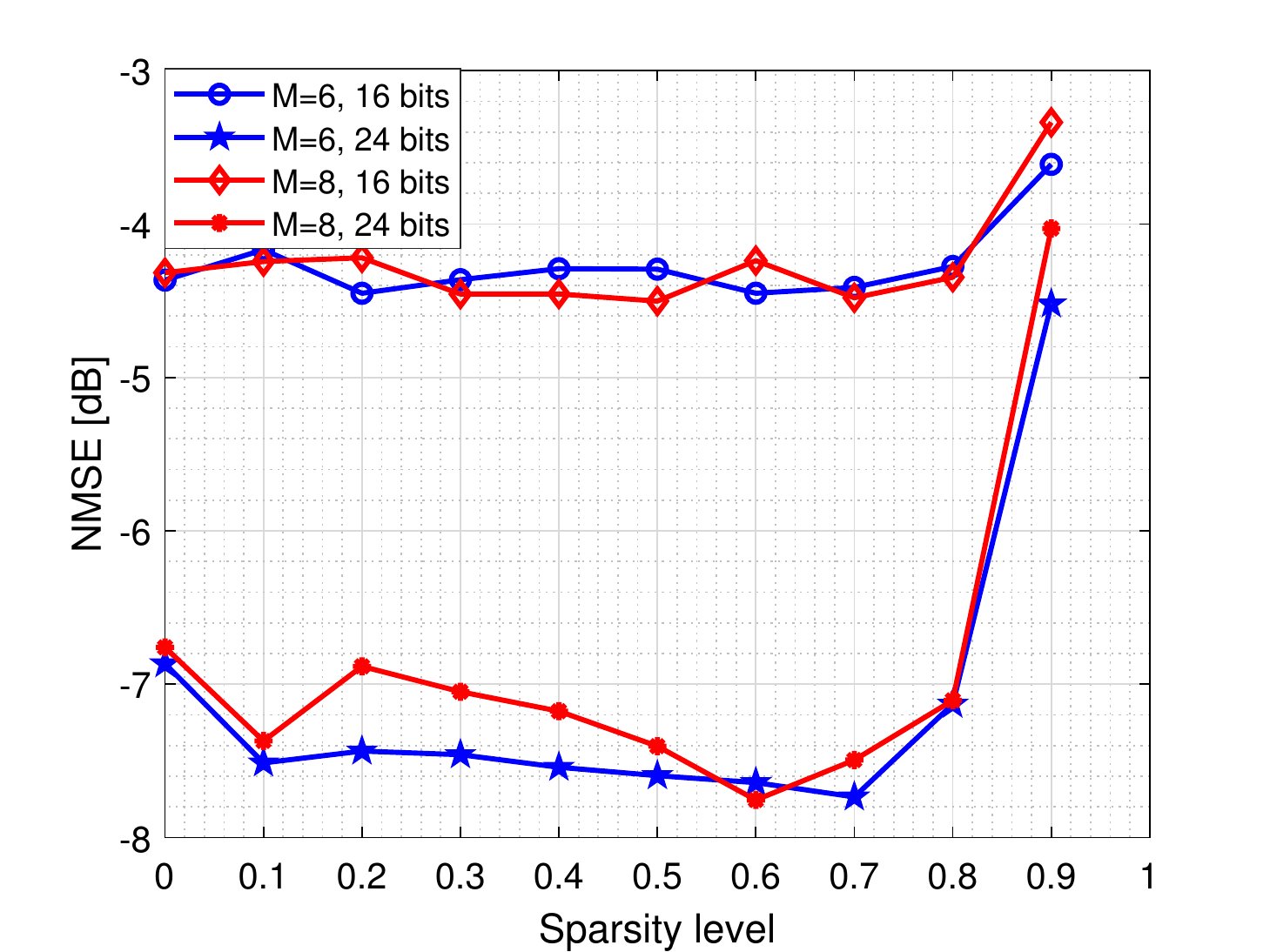}
    \caption{\label{prune} NMSE performance versus sparsity level of the CAnet-J framework with different pilot lengths and feedback bit numbers. The training and test SNRs are both set to 10 dB.}  
\end{figure*}

In this part, we prune the proposed CAnet-J by using the mechanism recommended by \cite{zhu2017prune}, and the initial sparsity value $s_{\rm i}$ is set to zero.
From (\ref{numberP}), (\ref{numberF}), the FLOP number is nearly proportional to the parameter number.
Therefore, in this part, we evaluate only the sparsity level, which can describe the parameter number.
Fig. \ref{prune} shows the NMSE performance versus sparsity level of the CAnet-J framework with different pilot lengths and feedback bit numbers. The training and test SNRs are both set as 10 dB.
If the sparsity level is zero, then the NNs are not pruned.
If the sparsity level is 0.9, then only 10\% of the NN weights in the CAnet-J are reserved and the others are dropped. 
From the figure, when the sparsity level is below 0.8, the NMSE performance fluctuates and sometimes even outperforms the original NNs that are not pruned.
Given that the NNs are unavoidably redundant, a performance gap exists between the training and test datasets, which can be regarded as overfitting.
Many redundant connections are dropped through NN weight pruning, thereby increasing the generality of the NNs \cite{9136588,zhu2017prune}.
The simulation results also indicate that NN weight pruning can decrease the complexity by approximately 80\% without any performance drop.

\section{Conclusion}
\label{s7}

In this paper, we have proposed a DL-based uplink-aided downlink channel acquisition framework, CAnet, for FDD massive MIMO systems.
Unlike the most existing works, which mainly focus on the feedback module, we have considered the entire downlink CSI acquisition process.
The key idea of this work is to improve the accuracy of the downlink channel obtained by the BS via exploiting the correlation among bidirectional channels.
First, we have designed an adaptive pilot, which depends on the magnitude of the uplink channel.
Then, to avoid the bit allocation problem between the feedback of the downlink channel phase and magnitude, we feed back the uplink in the form of a complex number and embed the uplink magnitude information at the decoder. 
Lastly, we have combined the above modules and proposed two entire downlink channel acquisition frameworks, i.e., CAnet-J and CAnet-S.
We have found that the joint pilot design, channel estimation, and feedback can introduce additional gains, which can provide a guideline for future research.
For instance, CAnet-J with $M=6$ pilots and $N_{\rm bits}=20$ feedback bits outperforms CAnet-S with $M=8$ pilots and $N_{\rm bits}=24$ feedback bits.
Considering the high complexity of NN-based CSI acquisition methods, we have applied NN weight pruning to the proposed CAnet-J, thereby reducing the NN complexity by approximately 80\%.

\ifCLASSOPTIONcaptionsoff
  \newpage
\fi

\bibliographystyle{IEEEtran}
\bibliography{IEEEabrv,reference}
\end{document}